\documentclass[accepted]{uai2021} 

\usepackage[american]{babel}
\usepackage{mathtools} 

\usepackage{hyperref}

\usepackage{natbib} 
\usepackage{booktabs} 
\usepackage{tikz} 
\usepackage[utf8]{inputenc} 
\usepackage[T1]{fontenc}    
\hypersetup{colorlinks,linkcolor={blue},citecolor={red},urlcolor={blue}}
\usepackage{url}            
\usepackage{booktabs}       
\usepackage{amsfonts}       
\usepackage{nicefrac}       
\usepackage{microtype}      
\usepackage{lipsum}

\usepackage{notes2bib}
\usepackage{amssymb}
\usepackage[linesnumbered,ruled,boxed]{algorithm2e}
\usepackage{bm}
\usepackage{array}
\usepackage{pbox}
\usepackage{xcolor}

\usepackage{graphicx}
\usepackage{subcaption}
\usepackage{float} 

\SetKwFor{ForAll}{for all}{do}{end}
\SetKwRepeat{Repeat}{repeat}{end}

\newcommand*{\vertbar}{\rule[-1ex]{0.5pt}{2.5ex}}
\newcommand*{\horzbar}{\rule[.5ex]{2.5ex}{0.5pt}}

\let\oldnl\nl
\newcommand{\nonl}{\renewcommand{\nl}{\let\nl\oldnl}}

\newtheorem{theorem}{Theorem}
\newtheorem{corollary}{Corollary}

\newtheorem{lemma}{Lemma}

\newcommand*{\qed}{\hfill\ensuremath{\square}}%
\newcommand{\indep}{\perp \!\!\! \perp}

\def\x{{\mathbf x}}
\def\y{{\mathbf y}}
\def\z{{\mathbf z}}

\DeclareMathOperator*{\argmin}{arg\,min}

\newcommand{\lambdab}{\boldsymbol{\lambda}}
\newcommand{\pib}{\boldsymbol{\pi}}
\newcommand{\mub}{\boldsymbol{\mu}}

\newcommand{\E}{\mathbb{E}}
\newcommand{\zest}{\widehat Z}

\def\mcA{\mathcal{A}}

\title{Optimized Auxiliary Particle Filters: \\  
adapting mixture proposals via 
convex optimization}

\author[1]{Nicola Branchini}
\author[1]{Víctor Elvira}
\affil[1]{%
    University of Edinburgh
}

\begin{document}
\maketitle

\begin{abstract}
Auxiliary particle filters (APFs) are a class of sequential Monte Carlo (SMC) methods for Bayesian inference in state-space models. In their original derivation, APFs operate in an extended state space using an auxiliary variable to improve inference. In this work, we propose \emph{optimized auxiliary particle filters}, a framework where the traditional APF auxiliary variables are interpreted as weights in a importance sampling mixture proposal. Under this interpretation, we devise a mechanism for proposing the mixture weights that is inspired by recent advances in multiple and adaptive importance sampling. In particular, we propose to select the mixture weights by formulating a convex optimization problem, with the aim of approximating the filtering posterior at each timestep. Further, we propose a weighting scheme that generalizes previous results on the APF (Pitt et al. 2012), proving unbiasedness and consistency of our estimators. Our framework demonstrates significantly improved estimates on a range of metrics  compared to state-of-the-art particle filters at similar computational complexity in challenging and widely used dynamical models. 
\end{abstract}

\section{Introduction}\label{sec:intro}
State-space models (SSMs) allow a mathematical description of complex dynamical systems which are very relevant in computational statistics, machine learning and signal processing, among many other fields \citep{sarkka2013bayesian}. Particle filters (PF) or sequential Monte Carlo methods (SMC) are the \emph{de facto} family of algorithms to perform inference tasks in virtually any SSM, e.g., filtering, prediction, or parameter estimation
\citep{doucet2001introduction}. PFs have been used for solving complex real-world problems in robotics  \citep{thrun2002particle}, object tracking \citep{vlassis2002auxiliary,wardhana2013mobile} and image processing \citep{nummiaro2003adaptive}.  
PFs are also used for problems beyond the classical SSM setting. For instance, they have been recently applied in reinforcement learning \citep{DBLP:conf/iclr/MaddisonLTHDMT17,ijcai2020wang,piche2018probabilistic}, generative modelling
\citep{lawson2018twisted,DBLP:conf/iclr/LeIRJW18},
and more generally for approximate Bayesian inference in large probabilistic models \citep{gu2015neural,pmlr-v84-naesseth18a,DBLP:conf/aaai/MaKHL20}. 
PFs are Monte Carlo methods that approximate probability density functions (pdfs)
of interest with $M$ particles. The \emph{bootstrap PF} (BPF) \citep{gordon1993novel} is the most popular algorithm, because of its simplicity and reasonable performance in several settings. However, alternatives are needed for challenging applications that require models with complex posterior distributions. Most notably, the \emph{auxiliary PF} (APF) \citep{pitt1999filtering} was designed to make use of the observation before the simulation of the particles. \\
 In this paper, we develop a framework named \emph{optimized} APF (OAPF) for accurate inference in SSMs. The OAPF framework implements a mixture proposal sampling and an associated weighting scheme at each time step within the PF, allowing for variance reduction in the importance weights, the key aim in SMC methods \citep{doucet2009tutorial}. 

  The structure of the paper is as follows. In Section~\ref{sec:back}, we review SSMs and give a brief overview on PFs.  
 In Section \ref{sec:meth}, we derive our OAPF framework, discussing the design choices and providing a theoretical analysis of its estimators. 
 In Section \ref{sec:exp}, we show improved results against common particle filters and the recent improved APF \citep{elvira2018search} in challenging and widely used nonlinear state-space models such as a stochastic Lorenz 63 model and a multivariate stochastic volatility model. We conclude the paper in Section \ref{sec:conclusion} with some final remarks.
 
\paragraph{Contributions.} (1) We develop the optimized auxiliary particle filter (OAPF) framework, which encompasses other particle filters as special cases and allows the development of new algorithms with improved estimators. Our framework has a flexible mixture proposal distribution which appears in the importance weights, provably reducing their variance.

(2) We prove that the resulting marginal likelihood estimators are unbiased and consistent, generalizing the APF estimator in \citep{pitt2012some}. 

(3) We propose strategies to select kernels and mixture weights in the proposal. The mixture weights are optimized by matching proposal and posterior at a set of relevant points. Crucially, this allows us to find mixture weights as a solution to a \emph{convex} optimization problem. Therefore, our strategy allows for optimizing the proposal in very  generic models (transition and observation pdfs), while avoiding black-box non-convex optimization methods that are common in for instance in variational inference \citep{archer2015black,dieng2017variational}. Further, we allow for a flexible choice of the number of kernels, detaching this choice from the number of particles unlike previous works (see for instance \citep{elvira2019elucidating}). 

(4) We propose specific implementations of our framework and show their effectiveness with widely used state-space models. We compare to BPF, APF and to the improved APF  (IAPF) \citep{elvira2018search}, a recent algorithm which provides the state-of-the-art in terms of importance weight variance. We show evidence for better estimates in OAPF with similar computational complexity. 

\section{Background}\label{sec:back}
\subsection{State-Space Models and Particle Filtering} 
State-space models (SSM) describe the temporal evolution of a system in a probabilistic manner. They are composed of a stochastic discrete-time Markovian process of a (potentially multivariate) \emph{hidden state} $\left \{ \x_t \right \}_{t \geq 1} $, which can only be observed via corresponding noisy measurements  $\left \{ \y_t \right \}_{t \geq 1} $. SSMs are fully specified by a prior probability density function (pdf), $p(\x_0)$, and by the \emph{transition} and \emph{observation} kernels, $f(\x_t |  \x_{t-1})$ and $g(\y_t |  \x_t)$, respectively, defined for $t \geq 1$. 
In these models, the \emph{filtering} task consists in the sequential estimation of the filtering density $p(\x_t |  \y_{1:t})$, as well as expectations of the form $I(h_{t} ) = \mathbb{E}_{p(\x_{t} |  \y_{1: t})}[h_{t}(\x_{t})] = \int h_{t}(\x_{t}) p(\x_{t} | \y_{1: t}) d \x_{t} ,$
for (integrable) functions of interest. 
For most SSMs of interests, the filtering pdf is intractable and 
one needs to resort to approximate inference. In this context, particle filters (PFs) are the most popular inferential methods, approximating the filtering pdf with a set of random particles (Monte Carlo samples).
PFs are a sequential implementation of importance sampling (IS), generating at each time step $M$ particles $\{\x_{t}^{(m)} \}_{m=1}^{M}$  from a proposal pdf $q(\x_t)$ and assigning them normalized importance weights $w_{t}^{(m)}$. The unnormalized importance weights
can be computed by updating the previous weights as
\begin{equation}\label{eq:smcweights}
  \widetilde{w}_{t}^{(m)} = w_{t-1}^{(m)} \frac{g(\y_t |  \x_{t}^{(m)} ) f(\x_{t}^{(m)} |  \x_{t-1}^{(m)})}{q (\x_{t}^{(m)} |  \y_{t}, \x_{t-1}^{(m)})}  ,
\end{equation}
which can be derived by factorizing a joint proposal $q(\x_{1:t-1} | \y_{1:t}) q(\x_t | \y_{t}, \x_{t-1})$ and targeting joint posterior $p(\x_{1:t} | \y_{1:t})$ \citep{sarkka2013bayesian}. Therefore, a particle filter maintains a set of normalized weights and particles $\{w_{t}^{(m)}, \x_{t}^{(m)}\}_{m=1}^{M}$ as a representation of the filtering pdf, updating weights at each time step with as in Eq. \eqref{eq:smcweights}. The most popular choice for $q(\x_{t} |  \y_{t}, \x_{t-1})$ is $f(\x_t |  \x_{t-1})$ and leads to the bootstrap particle filter (BPF) \citep{gordon1993novel}. The advantage of this choice is that the weights in \eqref{eq:smcweights} simply become $w_{t-1}^{(m)}  g(\y_t |  \x_{t}^{(m)})$. In practice, particle filters suffer from the \emph{weight degeneracy} effect \citep{sarkka2013bayesian}, consisting on few normalized weights taking all probability mass (i.e., the posterior is approximated with very few samples). In the BPF, a resampling step is introduced to mitigate this effect. In some implementations, the resampling step is performed only when the effective sample size $\text{ESS} = \frac{1}{\sum_{m=1}^M \left( w_t^{(m)} \right)^2}$ is below some threshold \citep{doucet2001introduction,doucet2009tutorial,sarkka2013bayesian}.   %

\subsection{ Auxiliary Particle Filters}
Auxiliary PFs (APFs) were introduced to alleviate some of the limitations of existing PF methods \citep{pitt1999filtering}. For instance, it is well known that informative likelihoods often impact negatively the ability of the standard BPF to reconstruct the filtering pdf \citep{doucet2009tutorial,johansen2008note,whiteley2011auxiliary}.\footnote{Informally, an informative likelihood refers to a peaky likelihood that heavily influences the shape of the posterior.} Intuitively, the reason is that the resampling step at the end of the recursion at time $t-1$ does not take into account the new observation $\y_t$. 
In the standard APF, the resampling step at $t-1$ is delayed until the new observation $\y_t$ is available. Then the resampling is performed with modified unnormalized weights
\begin{equation}\label{eq:concreteapfweights}
    \widetilde{\lambda}_{t}^{(m)} =  w_{t-1}^{(m)}  g(\y_t |  \mub_{t}^{(m)}) , \qquad \lambda_{t}^{(m)} = \frac{\widetilde{\lambda}_{t}^{(m)}}{\sum_{i=1}^{M}\widetilde{\lambda}_{t}^{(i)}}, 
\end{equation}
where $ \mub_{t}^{(m)} = \mathbb{E}_{f(\x_t | \x_{t-1}^{(m)})}[\x_t]$. Then the particles are propagated using the transition kernel $f(\x_t | \x_{t-1})$ as in BPF. Finally, the importance weights are chosen as
\begin{equation}\label{eq_apf_weights}
   \widetilde{w}_{t-1}^{(m)} =  \frac{g(\y_t | \x_{t}^{(m)})}{g(\y_{t} | \mub_{t}^{(i^{(m)})})} ,
\end{equation}
where $i^{(m)}$ denotes the index of the ancestor that generates the $m$-th resampled particle. Intuitively, this can be seen as scaling down the BPF weights, taking into account that particles have been already resampled in large number in regions of high likelihood. A different interpretation of APFs \footnote{Note that it is also possible to refer to as APF to a generic PF with $\lambda_t$ being a free choice. Eq. \eqref{eq:concreteapfweights}  is an approximation to $p(\mathbf{y}_t | \mathbf{x}_t)$} is possible from the multiple importance sampling (MIS) perspective \citep{elvira2019elucidating}. Note that MIS refers to the different sampling and weighting schemes that are possible in the presence of multiple proposals in IS \citep{veach1995optimally,elvira2019generalized}. In this perspective, a resampling step followed by a propagation step is considered to be simply a single sampling step from a mixture pdf. The improved APF (IAPF)~\citep{elvira2018search} exploits the MIS interpretation so that the weight of the $m$-th proposal, $\lambda_{t}^{(m)}$, depends on the location of other particles $j \neq m$. The MIS perspective is related to auxiliary marginal particle filters (AMPF) \citep{Klaas05}, where a similar importance weight is derived, but $\lambda_{t}^{(m)}$ is chosen as in APF. It is worth noting that \cite[Chapter~4,Section~3.2]{fearnhead1998} earlier analysed the basic idea behind AMPF.
 The AMPF interprets that the inference is performed in the marginal space of $\x_t$ (marginalizing the auxiliary variable), which guarantees to reduce (in a non-strict sense) the variance of the importance weights (it is a Rao-Blackwellization that can be proved by the variance decomposition lemma).
\paragraph{Optimality criteria for APF.} A version of the APF known as the \emph{fully adapted} APF (FA-APF) is considered to implement a locally optimal choice. Its implementation requires the computation of the (generally) intractable distributions $p(\x_t | \x_{t-1}, \y_{t})$ and $p(\x_{t-1} | \y_{1:t})$. While often presented as \emph{the optimal choice}, in \citet{doucet2009tutorial} the FA-APF is shown to provide worse estimators than the BPF in one example. The reason is that  FA-APF minimizes the variance considering only one step ahead, as explained thoroughly in \citep{chopin2020introduction}. Therefore, the intractability of FA-APF as well as its only relative optimality motivates the search for better PFs.

\section{Optimized Auxiliary Particle Filters}\label{sec:meth}

 \subsection{The OAPF Framework}
In this section, we present our {new} framework for \emph{optimized auxiliary particle filters} (OAPFs). The OAPF framework extends the MIS perspective, considering a generic mixture as proposal where all samples are (independently) simulated. We consider the generic mixture proposal at each $t$ as 
 %
\begin{equation}\label{eq:proposal}
   \psi_{t}(\x_t) = \sum_{k=1}^{K} \lambda_{t}^{(k)}  q_{t}^{(k)}(\x_{t}),
\end{equation}  
with associated mixture weights $\lambda_{t}^{(k)}$. To the best of our knowledge, the OAPF is the first method to detach the choice of $K$ from the number of samples $M$ (i.e., $K\neq M$ in the general case). 

\begin{algorithm}[ht]
\KwIn{prior, transition, and observation pdfs $p(\x_0), f(\x_t |  \x_{t-1}), g(\y_t |  \x_t)$, and sequence of observations $\y_{1:T}$}
\KwOut{set of weighted samples for each time step $\{\x_{t}^{(m)}, \widetilde w_{t}^{(m)}  \}_{m=1,t=1}^{M,T}$}
Draw $M$ samples from prior: $\x_{0}^{(m)} \sim p(\x_0)$ and set $w_{0}^{(m)} = 1/M$\;
\For{$t = 1,\dots, T $}{
\textit{(a) optimization step}: optimize the mixture proposal by selecting $K$ kernels $q_t^{(k)}$ and choosing their associated mixture weight $\lambda_t^{(k)}$ that compose the mixture proposal $\psi_t$ (see Section \ref{sec_opt}) 

  \textit{(b) sampling step}: simulate $M$ particles $\x_{t}^{(m)}$ from the proposal as
  \begin{equation}
    \x_{t}^{(m)} \sim \psi_t(\x_t)
  \end{equation}
  
\textit{(c) weighting step}: calculate new importance weights as: 
        \begin{equation}\label{eq_weights}
            \widetilde{w}_{t}^{(m)} = \frac{g(\y_t |  \x_{t}^{(m)}) \sum_{i=1}^{M} w_{t-1}^{(i)} f(\x_{t}^{(m)} |  \x_{t-1}^{(i)}) }{\sum_{k=1}^{K} \lambda_{t}^{(k)} q_{t}^{(k)}(\x_{t}^{(m)})} 
        \end{equation} \\
}
 \caption{Optimized Auxiliary Particle Filter}
 \label{alg_oapf}
\end{algorithm}

The  OAPF framework is described in Algorithm \ref{alg_oapf}. The method starts by simulating $M$ samples from the prior pdf, and then at each time $t$, it consists of the three following stages: (a) optimization, (b) sampling, and (c) weighting steps. Note that this structure keeps also some ties with adaptive IS (AIS) algorithms. In particular, the optimization step can be seen as an adaptive procedure of the mixture proposal with one iteration (see \citep{bugallo2017adaptive} for more details).  
First, the optimization step adapts the mixture proposal of Eq. \eqref{eq:proposal}. This procedure is discussed in detail in the next Section. Second, the new $M$ particles are simulated from the mixture proposal. Third, the importance weights are calculated as in Eq. \eqref{eq_weights}. 
It is worth remarking that the numerator does not evaluate the true filtering pdf but only an (unnormalized) approximation. However, the importance weights are still \emph{proper} (\citep{Liu04b}), as we show below.  

\subsection{OAPF importance weights}
The importance weights play a crucial role both in the estimators of generic moments of the approximate distributions and also in the behavior of the PF for the next time step. Hence, reducing the variance of the importance weights is the ultimate goal in PF. Since this variance depends on the discrepancy between the proposal and target pdfs \citep{ryu2014adaptive} the benefit of considering a mixture proposal in Eq. \ref{eq:proposal} and for the importance weights in Eq. \eqref{eq_weights} is twofold. First, mixtures are a flexible way to approximate a large collection of pdfs. Second, while PFs work implicitly with mixture proposal, only few works use them in the denominator of the importance weights \citep{Klaas05,elvira2018search,elvira2019elucidating}. Moreover, to the best of our knowledge these works did not extend the consistency results for the APF (Pitt 2012) to this importance weight. Placing the whole mixture in the denominator, as we do in OAPF, is known to reduce variance in MIS \citep{elvira2019generalized}, even yielding zero-variance weights in the case of perfect matching between the mixture proposal and target pdfs. 

In OAPF, the standard IS estimators can be built. More precisely, moments of the filtering pdf can be approximated by the self-normalized IS (SNIS) estimator as
\begin{align}\label{eq_snis}
\widehat I\left(h_{t}\right) = \sum_{m=1}^M w_t^{(m)}h_t(\x_t^{(m)}),
\end{align}
where $w_t^{(m)} = \frac{\widetilde w_t^{(m)} }{\sum_{j=1}^M \widetilde w_t^{(j)}}$ are the normalized weights. Finally, the weights of OAPF can be used to build an unbiased estimator of $p(\y_{1:t})$, which is crucial for many statistical tasks such as model selection \citep{luengo2020survey}. 
We build the OAPF estimator as:
   \begin{equation}\label{eq:jointconst}
     \widehat{p}(\y_{1:T}) = \widehat{p}(\y_{1}) \prod_{t=2}^{T} \widehat{p}(\y_t | \y_{1:t-1}),
 \end{equation}
where $\widehat{p}(\y_t | \y_{1:t-1}) = \frac{1}{M}  \sum_{m=1}^{M} \widetilde{w}_{t}^{(m)}$. 
The functional form of the OAPF estimator is similar to other PFs and can be justified by standard IS arguments, but the computation of the importance weights $\widetilde w_t^{(m)}$ differs from other methods as discussed above. In the following, we prove that the estimator $\widehat{p}(\y_{1:T})$ is unbiased and consistent, which turns the SNIS estimator of Eq. \eqref{eq_snis} consistent. 

\begin{theorem}
For any set of mixture proposals $\{\psi_t(\x_t)\}_{t=1}^T$ fulfilling standard regularity conditions in IS, the normalizing constant estimator in Eq.~\eqref{eq:jointconst} is unbiased and consistent, i.e., $\mathbb{E}[\widehat{p}(\y_{1:T})] = p(\y_{1:T})$ and $\lim_{M\to\infty} \widehat{p}(\y_{1:T}) = p(\y_{1:T})$ a.s. for any $T\in\mathbb{R}^+$.
\end{theorem}
\noindent \emph{Proof:} The proof is presented in the supplementary material as well as a description of the regularity conditions. \qed 

Note that the consistency of the SNIS estimator in Eq. \eqref{eq_snis} is also guaranteed by standard IS arguments (we complete this discussion in the supplement).
 
 Finally, note that the minimization of the variance of the normalizing constant is equivalent to minimizing the variance of the importance weights $\widetilde w_{t}$ \citep{doucet2009tutorial}. The OAPF explicitly aims at reducing this variance by minimizing the mismatch between the target pdf and the mixture proposal. In the supplement, we also present a proof showing that the variance the OAPF weights in Eq. \eqref{eq_weights} is always less than those of APF in Eq. \eqref{eq_apf_weights}, when $K=M$ and the mixture weights are the same. 
 
 \subsection{Optimization of the Mixture Weights}
\label{sec_opt}
In this section we discuss an approach to select the weights of the mixture proposal $\psi_t$. The ultimate goal is to select them so that the proposal is a good approximation of the approximate filtering posterior. To achieve this, we impose these two distributions to be pointwise close at a set of $E$ \emph{evaluation points} $\{\z_t^{(e)} \}_{e=1}^{E}$. We will show that this approach is flexible and brings several advantages. For simplicity, we start considering the case with $K=E$, where the evaluation points are the centers of the $K$ kernels $q^{(k)}$ in the proposal \eqref{eq:proposal}, i.e., $\{\z_t^{(e)} \}_{e=1}^{E} = \{ \mub_{t}^{(k)} \}_{k=1}^{K}$. More precisely, the $K$ kernels could be chosen as a subset $K$ elements from the set of $M$ transition kernels (from the previous $M$ particles). Note that APF and improved APF \citep{elvira2019elucidating} also use the center of the transition kernels. Alternatively, we could also use ``optimal'' SMC kernels \citep{doucet2009tutorial} which are defined via the intractable function $p(\x_t | \x_{t-1}, \y_t)$. Our framework allows for generic choices so these restrictions are not necessary. We continue this section in a generic setting, expanding the discussion on how many kernels and evaluation points to choose in Section ~\ref{sec:time}.

Now that we have fixed the $K$ mixture kernels and the $E$ evaluation points, we can satisfy the condition previously mentioned and build a \emph{linear} system of $E$ equations as:
   \begin{align}\label{eq:constrgeneralized}
  &\sum_{k=1}^{K} \lambda_{t}^{(k)}  q_{t}^{(k)}(\z_{t}^{(e)}) = \nonumber \\ &g(\y_t |  \z_{t}^{(e)}) \sum_{m=1}^{M} w_{t-1}^{(m)} f(\z_{t}^{(e)} |  \x_{t-1}^{(m)}) , ~~ e=1,\dots,E ,
\end{align}
where the $K$ mixture weights $\lambda_{t}^{(k)}$ are unknown at each time $t$. For a unique solution to exist is necessary that $K=E$, but in general we do not need to restrict to this case. Below, we show how to turn this problem into a (constrained) \emph{convex} optimization problem.
Let us define the the vectors $ \lambdab =  (\lambda_{t}^{(1)},\dots,\lambda_{t}^{(K)}  )^\top ,   \mathbf{w} =   (w_{t-1}^{(1)},\dots w_{t-1}^{(M)})^\top,  \mathbf{f}^{(e)} = (f(\z_{t}^{(e)} |  \x_{t-1}^{(1)}),\dots, f(\z_{t}^{(e)} |  \x_{t-1}^{(M)}))^\top, $ and  $ \mathbf{q}^{(e)}  = (q_{t}^{(1)}(\z_{t}^{(e)}),\dots, q_{t}^{(K)}(\z_{t}^{(e)} ))^\top$.
Then, we can re-write Eq. \eqref{eq:constrgeneralized} as 
\begin{align}\label{eq:constrgeneralizedvector}
\mathbf{q}^{(e)^{\top} }\lambdab = \underbrace{g(\y_t |  \z_{t}^{(e)}) \odot \mathbf{w}^\top  \mathbf{f}^{(e)}}_{ \widetilde{\mathbf{\pib}}^{(e)} }, 
\end{align}
for $e=1,\dots,E$, where $\odot$ is elementwise multiplication and defining additionally the right-hand side to be $\widetilde{\mathbf{\pib}}^{(e)}$. \\
More compactly, Eq. \eqref{eq:constrgeneralizedvector} can be re-expressed in matrix form as:
\begin{align}\label{eq:system}
\begin{split}
& \overbrace{\left[
  \begin{array}{ccc}
    \horzbar & \mathbf{q}^{(1)^\top }& \horzbar \\
    \vdots  &  \vdots & \vdots \\
    \horzbar &   \mathbf{q}^{(E)^\top } & \horzbar
  \end{array}
\right]}^{E \times K} 
\overbrace{\left[
  \begin{array}{c}
    \vertbar  \\
    \lambdab  \\
    \vertbar 
  \end{array}
\right]}^{K \times 1}
= \\
&\overbrace{\left[
  \begin{array}{ccc}
    \horzbar & g(\y_t |  \z_{t}^{(1)}) \odot \mathbf{f}^{(1)^\top }& \horzbar \\
    \vdots  &  \vdots & \vdots \\
    \horzbar &   g(\y_t |  \z_{t}^{(M)}) \odot \mathbf{f}^{(E)^\top } & \horzbar
  \end{array}
\right]}^{E \times M}
\overbrace{\left[
  \begin{array}{c}
    \vertbar  \\
    \mathbf{w}  \\
    \vertbar 
  \end{array}
\right]}^{M \times 1},  
\end{split}
\end{align}
defining $\mathbf{Q}$ as the $E \times K$ matrix on the left-hand side of \eqref{eq:system} and $\widetilde{\pib}$ as the resulting $E \times 1$ vector on the right-hand side. We now define a generic constrained optimization problem as for the mixture weights as:
\begin{equation}\label{eq:genericopt}
    \lambdab^{*} =  \argmin_{\lambdab} \mathcal{L} \left ( \mathbf{Q}\lambdab , \mathbf{\widetilde{\pib}} \right ) ,
\end{equation}
where $\mathcal{L}(\cdot)$ a generic loss function. The optimization will be constrained since $\lambdab$ will be used for resampling, and therefore needs to have non-negative elements.\footnote{The resulting values can be normalized afterwards so they parametrize the mixture proposal in Eq. \eqref{eq:proposal}.}
In in the next Section, we present a possible strategy to implement $\mathcal{L}(\cdot)$ and solve the optimization problem.
\paragraph{Optimization via Non-Negative Least Squares (NNLS)}\label{sec:optnnls}
The previous problem can be encoded as a non-negative least squares problem by taking the squared distance of the pdfs at the $E$ evaluation points $\{ \z_{t}^{(e)}\}_{e=1}^{E}$. Taking squared differences between left-hand side and right-hand side of \eqref{eq:system} leads to:
\begin{align*}
    \lambdab^{*} = \argmin_{\lambdab} \left\| \mathbf{Q}\lambdab - \mathbf{\widetilde{\pib}}  \right\|_{2}^{2} ~~~ \text{subject to}: \lambdab \in \mathbb{R}_{\geq 0}^{K}.
\end{align*}
This problem is a (constrained) quadratic program. Therefore, it is convex and the non-negativity constraints form a convex feasible set. When $\mathbf{Q}$ has full column rank, then there is a unique solution. Theoretical results on NNLS have shown that, especially when the dimension of $\mathbf{Q}$ is large (large $K$ and $E$ in our case), the solutions tend to be \emph{very sparse} \citep{slawski2013non,meinshausen2013sign}. The optimization problem can be solved by the widely used algorithm in \citep{lawson1995solving}, as well as by concurrent work on exact sparse NNLS \citep{nadisic2020exact} and even strong \emph{GPU accelerations} could be exploited \citep{luo2011efficient,kysenko2012gpu}. There are other possible choices to implement Eq. \eqref{eq:genericopt}. For instance, it is possible to formulate a linear program and solve it with the \emph{Simplex} algorithm. We tried this approach, but found that the stability of the algorithm may be endangered above $3$ to $5$ dimensions.

\subsection{Selection of Kernels, Evaluation Points, and Computational Complexity} \label{sec:time}

The generic OAPF framework also allows for the choice of the number and type of kernels. In our experiments, we choose the transition kernel for simplicity (as it is done in APF or BPF), but other choices are equally valid. Regarding the number of kernels $K$, the novel MIS perspective allows for an extra degree of freedom unlike in standard filters. In particular, we have found that it is in general possible to reduce $K$ dramatically w.r.t. $M$,  without a a loss of performance; howevere, the ESS as a degeneracy measure seems more sensitive to the decrease in $K$. Further work could develop a formal analysis to explain this behavior. Moreover, we are also able to reduce the number of target evaluations $E$ w.r.t $M$ at the optimization step (see more details in Section \ref{sec_opt}). Unlike in the APF where the number of pre-weights is necessarily $K=M$, in OAPF this is a choice. The reason is that the purpose of the evaluations is to evaluate the target pdf so the mixture proposal can place probability mass in relevant parts of the space. Therefore, the number of evaluation points ($E$) can be much smaller than the number of particles (more details are provided in the Supplement).

As a general guideline, our starting point is setting $K=E$ and choosing the evaluation points (deterministically) as the center of each $q_{t}^{(k)}$, retaining those associated with the $E$ \emph{greatest values of the approximate filtering pdf} in the RHS of Eq. ~\eqref{eq:constrgeneralized}: this simple scheme worked well in our experiments. Ultimately, we remark how more evaluation points can only improve the quality of the approximation to the filtering posterior. More strategies in evaluation points selection (number, deterministic selection vs sampling) could be explored in further work. 

The computational complexity of OAPF can be decomposed in weighting and optimization steps. In the former, the numerator of Eq. \eqref{eq_weights} has been shown to have very good approximations in Marginal PFs \citep{Klaas05} in time $\mathcal{O}(M \log M)$ using \emph{dual-tree} methods for weighted kernel density estimation problems. Further, due to the sparsity properties in the recovered $\boldsymbol{\lambda}$ (see Section \ref{sec:optnnls}), our method improves the effective runtime in the calculation of the importance weights. \\ The optimization step can be accelerated significantly by implementing standard methods of the rich literature on fast and accurate approximations to constrained least-squares problems  \citep{pilanci2016iterative}, or via direct application of Frank-Wolfe algorithms \citep{jaggi2013revisiting}.




\begin{figure*}[t]
    \centering
    \begin{subfigure}[t]{\columnwidth}
    \centering
        \makebox[\textwidth]{\includegraphics[width=0.90\columnwidth]{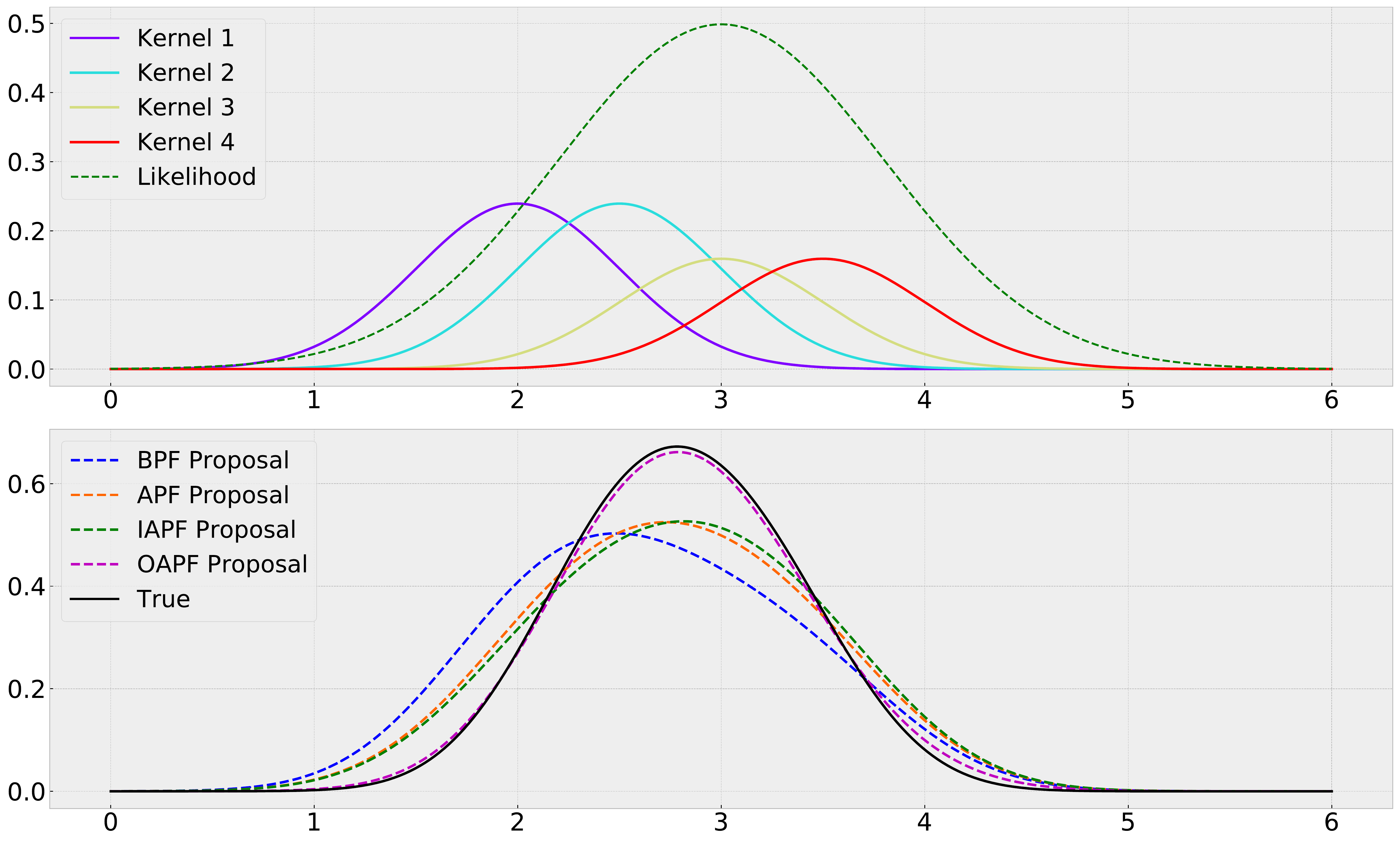}}
          \caption{In this first example we choose a unimodal posterior. OAPF substantially outperforms the other algorithms.}
          \label{fig:mog1}
    \end{subfigure} \hfill \begin{subfigure}[t]{\columnwidth}
    \centering
        \makebox[\textwidth]{\includegraphics[width=0.90\columnwidth]{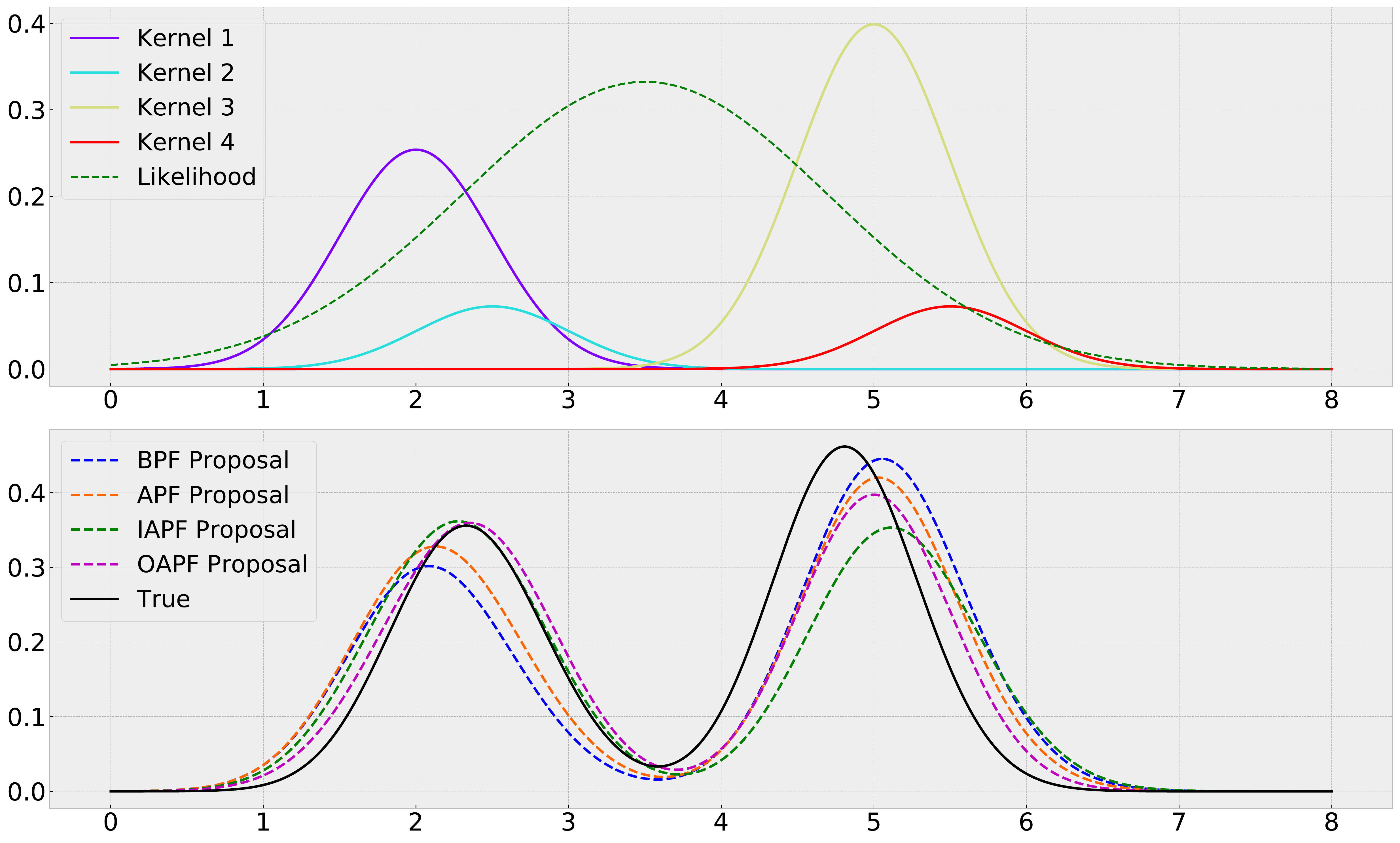}}
                  \caption{OAPF is the only algorithm who can match well \emph{both} modes simultaneously with this multimodal posterior.  }
          \label{fig:mog2}
    \end{subfigure}
\caption{\textbf{Experiment 1 (Toy Example.)} In this experiment we show that OAPF proposals are closer to true posteriors compared to its competitors. We calculated $\chi^2$-divergence for these examples in Table \ref{chisq}. Note that here OAPF uses transition kernels for the proposal and their centers as evaluation points. We provide all parameters for reproducibility in the supplement.}
\label{fig1}
\end{figure*}

\section{Related Work}

The OAPF follows a different approach w.r.t. most papers in the PF literature by interpreting the $M$ samples to be simulated from a mixture proposal with $K$ components. Moreover, unlike other popular PFs, we allow for a reduction of the number of the components, exploiting the sparsity behavior of the optimization algorithm. This perspective is connected to the auxiliary marginal PF (AMPF) \citep{Klaas05} and improved APF (IAPF) \citep{elvira2018search} algorithms, and is supported by recent advances in MIS \citep{elvira2019generalized} (see also the discussion of the variance reduction in \citep{Klaas05}), and it also links with the re-interpretation of BPF and APF \citep{elvira2019elucidating}. The selection of the mixture weights has connections with other works. For instance, a flexible framework named \emph{twisted} APFs is developed in \citep{guarniero2017iterated}, where APFs are interpreted as a special case of changing the distribution targeted in IS  (this interpretation appeared first in \citep{johansen2008note,doucet2009tutorial}. In this method, the computation is done in an offline fashion (see an extension of this line in \citep{heng2020controlled}). Resampling weights were also found in \citep{reich2013nonparametric} via a convex optimization problem derived via optimal transport arguments, which is however more computationally expensive than ours, and scales worse to higher dimensions. Further, our selection can be connected to \emph{black-box} importance sampling \citep{liu2017black}, which computes  IS weights\footnote{rather than simulation weights, which are specific to APF} in a static setting with convex optimization. In \citep{akyildiz2020nudging}, they develop a PF framework with a different approach, preemptively moving a subset of particles to a region of high likelihood with gradient methods. In \citep{cornebise2014adaptive}, the method adapts a mixture of kernels in a more generic setup (sequential Monte Carlo samplers), focusing in the choice of kernels. 
Further, the approach in \citep{kronander2014robust}  propagates  particles associated with kernels that are placed high likelihood regions, which could be combined in our framework.  

\section{Experiments}\label{sec:exp}

 We compare OAPF with BPF, APF as well as the recent improved APF (IAPF) \citep{elvira2018search}, which also uses a mixture in the denominator of the importance weights and can be seen as a special case in our framework. The IAPF strictly improves over APF and BPF in most settings \citep{elvira2018search}. Note that the simple BPF can sometimes perform unexpectedly well, as it is well known in the PF community. In the linear Gaussian model, we additionally compare with the fully adapted APF (FA-APF). We evaluate our framework in 4 sets of experiments.\\ Our aim is to show the benefits of OAPF in terms of variance of importance weights, which is crucial in particle filters: the weights are used not only for approximating integrals of interest but also for building better particle approximations in the next time steps. Therefore, we choose metrics that are directly connected to the variance of the importance weights: $\chi^2$-divergence between mixture proposal and filtering pdfs, error in the estimation of the posterior mean and marginal likelihood, and effective sample size (ESS) \citep{sarkka2013bayesian}. The setup of the experiments is as follows:

\begin{itemize}
    \item \textbf{Experiment 1: Toy example}. We show visually that the mixture proposal in OAPF reconstructs the posterior better than its competitors, both with unimodal and multimodal posteriors. Numerically we show an improved $\chi^2$-divergence between proposal and filtering pdfs, which directly translates into lower variance of importance weights.
    \item \textbf{Experiment 2: Linear dynamical model.} We exploit the closed-form solution of the linear dynamical model, perhaps the most known SSM and widely used for instance in object tracking \citep{sarkka2013bayesian}. This allows comparison with sampling from the Kalman Filter, as well as a closed form FA-APF.  We show that OAPF reaches better solutions with highly reduced runtime w.r.t IAPF thanks to our selection of $K$ and $E$.
    \item \textbf{Experiment 3: Stochastic Lorenz 63 model.} Transitioning to more challenging non-linear non-Gaussian models, we show an improved performance on discretized version of this popular chaotic dynamical system, which is used for instance in atmospheric models for weather forecasting \citep{ott2004local,yeong2020particle}.  We compare the PFs in terms of the ESS, which is widely used as a proxy for the weight variance.
    \item \textbf{Experiment 4: Stochastic volatility model.}  Finally, we perform inference for a multivariate stochastic volatility model used in related work on APFs \citep{guarniero2017iterated}. Here, as in Experiment 4, we look at ESS and show improved performance against all other algorithms.
\end{itemize}

We consider time-series with $T=100$ time steps, except otherwise stated. We let $d_{\x}$ be the dimension of the hidden state, i.e., $\x_t\in\mathbb{R}^{d_{\x}}$. Due to the curse of dimensionality, a general reduction in performance for all methods is expected as $d_{\x}$ grows. For linear dynamical models, we show improved estimates with significantly reduced runtime than all other algorithms, including IAPF. For the more challenging models where ground truth is not available, we achieve better ESS than the competitors. Note that in those experiments, we set $K=E=M$. However, due to the high sparsity of solutions in OAPF, our effective $K$ is much lower. For $M=1000$ particles we report an average of $88\%$ sparsity, while for $M=100$ an average of $65\%$. All averages and standard errors are obtained with $100$ independent Monte Carlo runs.\footnote{The code used in the experiments can be found at \url{https://github.com/nicola144/optimized_auxiliary_particle_filters}}

\begin{table}[h]
\caption{ \textbf{Experiment 1 (Toy example).} $\chi^2$-div. between filtering and mixture proposal pdfs in Figure \ref{fig1}.} \label{chisq}
\begin{center}
  \begin{tabular}{ccc}
Method  &  $\chi^2$-div.  (Fig. \ref{fig:mog1}) & $\chi^2$-div. (Fig. \ref{fig:mog2}) \\
\hline \\
BPF        & $0.1662$& $  0.2245$  \\
APF          & $0.0916 $ & $0.1633$  \\
IAPF             & $0.0870$ & $0.2402$ \\
OAPF             &  $\boldsymbol{0.0069 }$ & $\boldsymbol{0.0819} $
\end{tabular}
\end{center}
\end{table}


\begin{figure*}[t]
    \centering
    \begin{subfigure}[t]{\columnwidth}
    \centering
        \makebox[\textwidth]{\includegraphics[width=0.85\columnwidth]{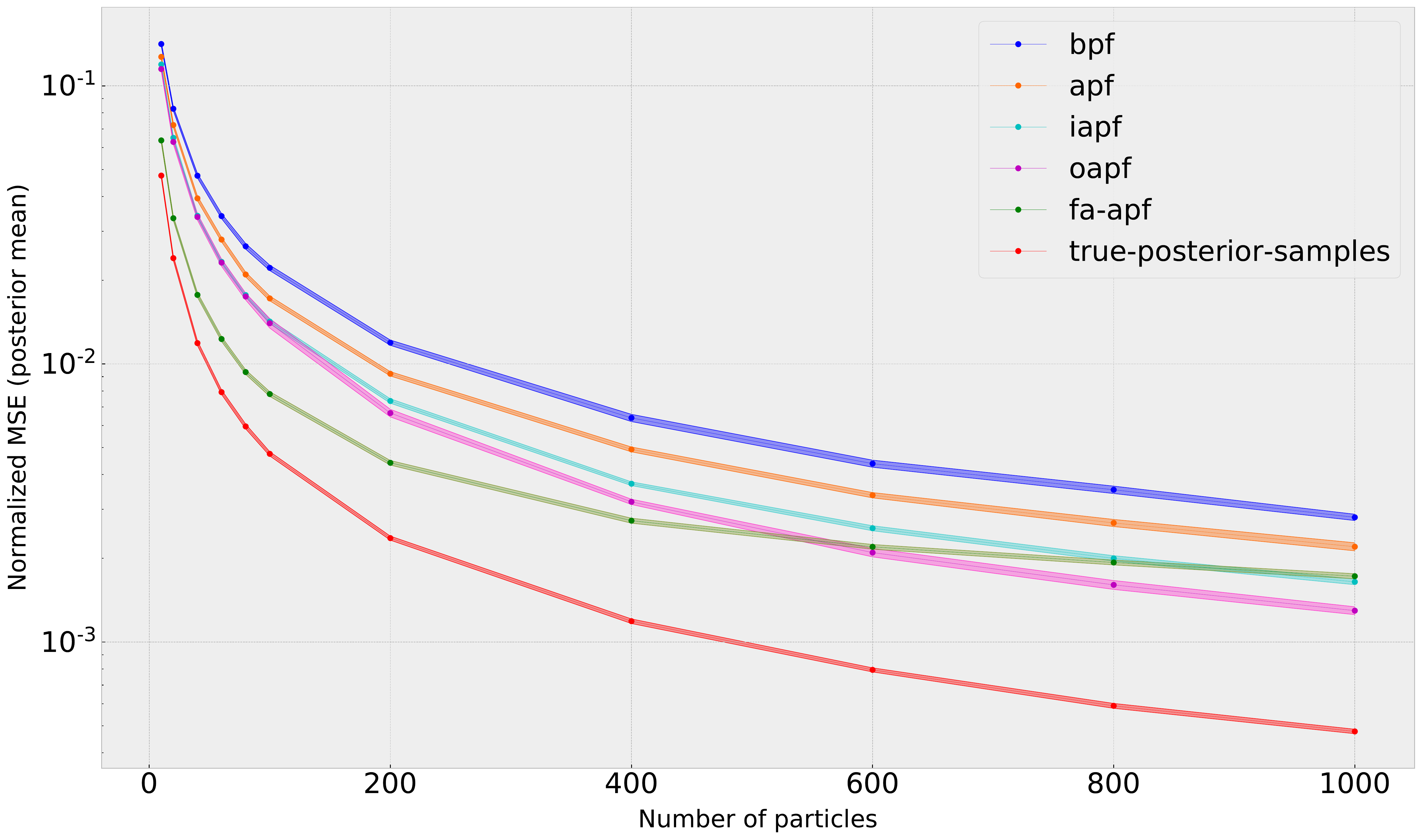}}
        \caption{ }
          \label{fig:ess1}
    \end{subfigure} \hfill \begin{subfigure}[t]{\columnwidth}
    \centering
        \makebox[\textwidth]{\includegraphics[width=0.85\columnwidth]{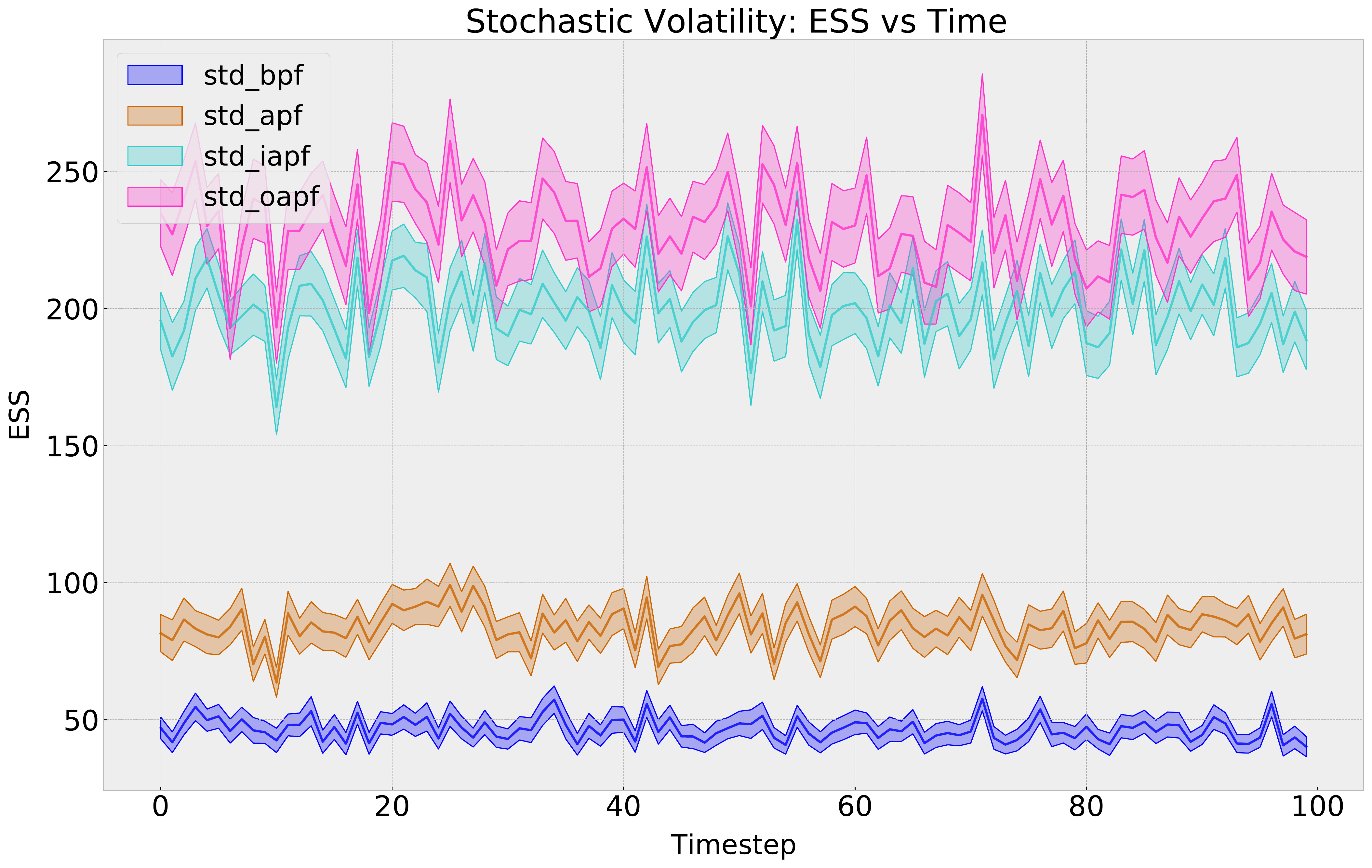}}
        \caption{ }
          \label{fig:ess2}
    \end{subfigure}
\caption{ (a) \textbf{Experiment 2 (Linear dynamical model).} NMSE to the true posterior mean as function of the number of particles, with $d_{\x} = 10$. Note that OAPF runs with $\boldsymbol{K=E=5}$ in all cases, selected by the strategy described in Section \ref{sec:time}. The FA-APF is only available analytically in this model, and the line true posterior samples (sampled from the Kalman filter) is presented as benchmark. (b) \textbf{Experiment 4 (Stochastic volatility).} ESS as a function of time with $d_{\x} = 10$, $M=1000$.}
\label{fig:ess}
\end{figure*}

\paragraph{1. Toy Example.}
The main goal of this toy example is to illustrate that the OAPF mixture proposal better reconstructs the filtering pdf. We also measure the $\chi^2$-divergence between both pdfs. We consider a single iteration of each PF algorithm and build artificial proposals by multiplying a mixture of $4$ Gaussians with a Gaussian likelihood. 
Results from the two experiments with the above setting are shown in Figure \ref{fig1}. We select the means of the transition kernel $f(\cdot)$ as evaluation points, and set $K=E=M=4$.  In Figure~\ref{fig:mog1}, we show the results for a unimodal posterior. This setting is advantageous for the IAPF, as transition kernels significantly overlap (see \citep{elvira2019elucidating} for more details). The likelihood is sufficiently informative, which explains why APF outperforms BPF. 
Figure  \ref{fig:mog2} shows a more complex multimodal posterior with a more diffused likelihood. 
Interestingly, we find that IAPF can perform even worse than APF, while our OAPF does not suffer from this issue. Table~\ref{chisq} quantifies (for both settings) the mismatch between mixture proposal and filtering pdfs in terms of $\chi^2$-divergence, confirming the visual analysis of Fig. \ref{fig:mog1}.
\paragraph{2. Linear Dynamical Model.}
The linear dynamical model is arguably the most popular SSM, routinely being the first choice to assess PFs. It has been applied in a wide range of applications (e.g., robotics \citep{sarkka2013bayesian}). This model is particularly useful for validating PFs since it is one of the few models admitting closed-form solutions of the filtering distribution and the normalizing constant (via the celebrated \emph{Kalman filter}). The defining transition and observation equations are standard (see e.g., \citep{sarkka2013bayesian}) and require mainly the selection of observation and transition covariances (more details in the supplement).
We tested the algorithms for different choices of model parameters, e.g.,  for high-variance observation noise or high-variance transition noise. 
In Figure~\ref{fig:ess1}, we calculate the normalized mean-squared error (NMSE) \footnote{We define NSME as mean-squared error divided by true value.} between the estimate of the posterior mean and the true value for $M \in \{ 10,20,40,60,80,100,200,400,600,800,1000 \}$ with $d_{\x} = 10$. Here, observation and transition covariances were set to $5  \mathbf{I}$ and $2.5 \mathbf{I}$ respectively; complementary results with other parameters, dimensions, as well as additional results on estimation of the normalizing constant are available in the supplement. The proposed OAPF outperforms all the competitors with a low $K=E=5$ in all settings, which is translated into large computational savings. Similar conclusions can be extracted for other choices of the model parameters. 




\paragraph{3. Stochastic Lorenz 63 Model.}
The Lorenz 63 is a \emph{chaotic} system, since slightly different initial conditions generate extremely different trajectories. Due to this difficulty, this model is often used to evaluate PFs \citep{akyildiz2020nudging}. We consider a discretized version of the state dynamics using an Euler-Maryuyama scheme and observations with additive noise. The hidden state is three dimensional $\x = [x^{[1]},x^{[2]},x^{[3]}]$ and the transition dynamics are defined by the differential equations:
\begin{align}
    &\mathrm{d}x^{[1]} = \sigma (x^{[2]} - x^{[1]}) \mathrm{d}\tau + \mathrm{d}w_{x^{[1]}} \\
    &\mathrm{d}x^{[2])} = (\rho x^{[1]} - x^{[3]} x^{[1]} - x^{[2]}) \mathrm{d}\tau + \mathrm{d}w_{x^{[2]}} \\
    &\mathrm{d}x^{[3]} = (x^{[1]} x^{[2]} - \beta x^{[3]}) \mathrm{d}\tau +  \mathrm{d}w_{x^{[3]}}
\end{align}
where $\tau$ denotes continuous time, $w_{x^{[1]}},w_{x^{[2]}},w_{x^{[3]}}$ are independent one-dimensional standard Wiener processes and $(\sigma,\rho,\beta)$ are parameters of the model. We use the an increment $\Delta t$ in the discretization, and \emph{partially} observe  the hidden state (only the first dimension) with scalar $y_{t} \sim \mathcal{N}_{y_t}(x^{(1)},\sigma_{y_{t}}^{2} = 1)$, using the standard values for $(\sigma,\rho,\beta)$ (see supplement). The results of averaged ESS with two different values of $\Delta t\in\{0.01,0.008\}$ are shown in Table~\ref{lorenz}. Note that even these small changes in $\Delta t$ cause very different trajectories, as we also show in the supplement.

\paragraph{4. Stochastic Volatility Model.}
We perform inference in a multivariate stochastic volatility model (SVM), a type of stochastic process where the variance is a latent variable that follows itself a stochastic process. These are extremely useful models to apply for many tasks in econometrics, e.g., for predicting the volatility of a heteroskedastic sequence such as returns on equity indices or currency exchanges. SVMs are often used to evaluate particle filters \citep{pitt1999filtering,Klaas05,guarniero2017iterated}. We employ the  
version in \citep{chib2009multivariate}, which is also used in related work on APFs \citep{guarniero2017iterated}. It is defined by the following pdfs: 
\begin{align}
    p(\x_0) &= \mathcal{N}_{\x_0} \left ( \mathbf{m}, \mathbf{U}_0 \right ), \\
    f(\x_t |  \x_{t-1}) &= \mathcal{N}_{\x_{t}} \left(\mathbf{m}+ \text{diag}(\mathbf{\phi})(\x_{t-1} - \mathbf{m}), \mathbf{U} \right),  \\
    g(\y_t |  \x_t ) &= \mathcal{N}_{ \y_t } \left ( \mathbf{0}, \exp \left ( \text{diag}(\x_{t}) \right ) \right ). 
\end{align}
 Table~\ref{stochvol} shows averaged ESS for $d_{\x} = (2,5,10)$.
The parameters for this experiment are set to $\mathbf{m} = \mathbf{0}, \mathbf{U}_0 = \mathbf{I}, \mathbf{U} =  \mathbf{I}, \mathbf{\phi} =  \mathbf{1} $. Figure \ref{fig:ess2}  shows the averaged ESS over time for $d_{\x}=10$ and same parameters, except $\mathbf{\phi} =  \frac{1}{2}\mathbf{1}$. For additional results, see supplementary.

 \begin{table}[t]
\caption{ \textbf{Experiment 3 (Lorenz)}. $T=1000$ timesteps, $M=100$ particles. Averaged ESS and standard errors.} \label{lorenz}
\begin{center}
   \begin{tabular}{ccc}
Method  & $\Delta t = 0.01$  &  $\Delta t = 0.008$   \\
\hline \\
BPF    &   $ 57.7 \pm 0.2 $  & $58.1 \pm 0.2$  \\
APF      & $ 55.1 \pm 0.2  $  &  $55.2 \pm 0.2$ \\
IAPF        &  $ 70.1 \pm 0.1  $ &  $71.0 \pm 0.1$ \\
OAPF         & $ \boldsymbol{76.7 \pm 0.1}  $ & $ \boldsymbol{76.4 \pm 0.1}$ \
\end{tabular}
\end{center}
\end{table}



\begin{table}[t]
\caption{ \textbf{Experiment 4 (Stochastic Volatility)}. Note that when $d_{\x}=10$ then $M=1000$, otherwise $M=100$.  Averaged ESS and standard errors.} \label{stochvol} 
\begin{center}
  \begin{tabular}{cccc}
Method  &  $d_{\x} = 2$ & $d_{\x} = 5$ & $d_{\x} = 10$\\
\hline \\
BPF        & $63.5 \pm 0.2 $& $ 33.5\pm 0.2$  & 108.7 $\pm$ 0.8 \\
APF          & $63.5 \pm 0.2 $ & $34.5 \pm 0.2$ &  107.2 $\pm$ 1.0  \\
IAPF             & $73.0 \pm 0.1$ & $44.9 \pm 0.2$ & $ 203.5 \pm 0.9 $\\
OAPF             &  $\boldsymbol{ 88.3 \pm 0.2 }$ & $\boldsymbol{63.5 \pm 0.2} $& $\boldsymbol{366.2 \pm 1.8}$
\end{tabular}
\end{center}
\end{table}




\section{Conclusions}
\label{sec:conclusion}
 In this paper we have proposed the OAPF, a flexible framework for particle filtering that uses a generic mixture distribution as a proposal and includes it in the importance weighting scheme. 
 The framework allows for the development of particle filters with improved performance, and we provide an explicit implementation. We have proved the unbiasedness of the OAPF marginal likelihood estimator for any mixture proposal that fulfills standard IS requirements. We also show the effectiveness of OAPF in reducing the varince of the IS estimators. In OAPF, we directly optimize the mixture proposal to the posterior in an online fashion, rather than making specific analytic choices of mixture weights like in AMPF or IAPF. Conversely to most other methods that optimize a proposal (e.g., variational inference), our optimization strategy is convex, directly addressing the ultimate goal of minimizing the variance of the importance weights. Therefore, OAPF can deal with any likelihood and transition models (that admit a density) without resorting to black-box, non-convex methods (see for instance \citep{archer2015black,dieng2017variational}). We have shown improved performance of the proposed implementation of the OAPF across a series of challenging state-space models and metrics, comparing with BPF, APF, and the competitive IAPF.
Finally, the flexibility and the strong theoretical guarantees of OAPF pave the way for new methodological advances within this framework.



\begin{acknowledgements} 
We would like to thank Theodoros Damoulas for initial comments on an earlier version of the paper, and for the support during the refinement of this work. 
\end{acknowledgements}

\clearpage

\onecolumn

\appendix
\section{Supplementary Material}

\subsection{Theoretical Properties of the OAPF Estimators}


The theoretical properties of the estimators OAPF are analized from the importance sampling perspective. In the case of the mixture proposals $\psi_t$, we assume that each time $t$, the support of $\psi_t$ is a superset of the support of $p(\x_t|\y_{1:t})$, i.e., that $\psi_t(\x_t)>0$ for all $\x_t$ where $p(\x_t|\y_{1:t})>0$.  Let us define the \emph{partial} normalizing constants as $Z_{t} \triangleq p(\y_{t}|\y_{1:t-1})$, the \emph{joint} normalizing constant as $Z_{1:t} \triangleq p(\y_{1:t})$, and also $Z_{t-h:t} \triangleq p(\y_{t-h:t}|\y_{1:t-h-1})$. In the OAPF framework, we can build estimator of those quantities, e.g., the partial estimator $\zest_\tau \triangleq \frac{1}{M}\sum_{m=1}^M \widetilde w_{\tau}^{(m)}$, the joint estimator $\zest_{1:t} = \prod_{\tau = 1}^t \zest_{\tau} $, and also the estimator $\zest_{t-h:t} = \prod_{\tau = t-h}^t \zest_{\tau} $, with  and the estimator $\zest_t \triangleq \frac{1}{M}\sum_{m=1}^M \widetilde w_{t}^{(m)}$. We also assume that the estimators of all the partial normalizing constants have finite variance (see for instance \citep{owen2013monte,elvira2019generalized}). We define the set of weighted samples at time $t$ as $\mcA_t \triangleq \{\x_t^{(m},\widetilde w_t^{(m} \}_{m=1}^M$. In order to avoid ambiguities when evaluating pdfs, we define the functions $g(\y_t|\x_{t})\triangleq p(\y_t|\x_{t})$, $g(\y_t|\x_{t-1})\triangleq p(\y_t|\x_{t-1})$, $g(\y_t,\x_t|\x_{t-1})\triangleq p(\y_t,\x_t|\x_{t-1})$ and $g(\y_{t-h:t},\x_t|\x_{t-1})\triangleq p(\y_{t-h:t},\x_t|\x_{t-1})$. 

In the following, we show that OAPF provides an unbiased estimator of the normalizing constant $p(\y_{1:t})$, which follows a proof by induction, in a similar spirit as in \citep{pitt2012some}, but with more generic results. In particular, here the (approximate) filtering distribution is the marginalized version of the one in \citep{pitt2012some} and is constituted by a mixture in the numerator of the importance weights (see \citep{Klaas05} for an explanation). In OAPF the proposal density can be any mixture $\psi_t(\x_t)$ fulfilling the standard regularity conditions described above, hence in the denominator of the importance weights, a second mixture appears.  Theorem \ref{th_unbiased_z} is here the main result, and is supported by Lemmas  \ref{lemma_6_pitt} and \ref{lemma_7pitt} which we present first.

\begin{lemma}\label{lemma_6_pitt}
We have that 
\begin{equation}
\E\Big[\zest_t|\mcA_{t-1} \Big]  = \sum_{m=1}^M w_{t-1}^{(m)} g(\y_t|\x_{t-1}^{(m)}).
\end{equation}
\end{lemma}
\noindent\emph{Proof:} 
\begin{align}
\E\Big[\zest_t|\mcA_{t-1} \Big]  &= \E\Big[\frac{1}{M}\sum_{m=1}^M   \widetilde w_{t}^{(m)} |\mcA_{t-1}\Big] \\
&= \E\Big[\frac{1}{M}\sum_{m=1}^M    \frac{g(\y_t|\x_t^{(m)})\sum_{j=1}^M w_{t-1}^{(j)}f(\x_t^{(m)}|\x_{t-1}^{(j)}) }{\psi(\x_t^{(m)})}|\mcA_{t-1} \Big] \\
&= \frac{1}{M}\sum_{m=1}^M    \E\Big[\frac{g(\y_t|\x_t^{(m)})\sum_{j=1}^M w_{t-1}^{(j)}f(\x_t^{(m)}|\x_{t-1}^{(j)}) }{\psi(\x_t^{(m)})} |\mcA_{t-1}\Big]. \label{step4} 
\end{align}
Now, since given $\mcA_{t-1}$ the particles at time $t$ are conditionally independent with pdf $\psi_t(\x_t)$, then we have that the integrals within \eqref{step4} are identical:
\begin{align}
\E \Big[\zest_t|\mcA_{t-1} \Big] &= \int \frac{g(\y_t|\x_t)\sum_{j=1}^M w_{t-1}^{(j)}f(\x_t|\x_{t-1}^{(j)}) }{\psi(\x_t)} \psi(\x_t)d\x_t  \\
&=  \int g(\y_t|\x_t)\sum_{j=1}^M w_{t-1}^{(j)}f(\x_t|\x_{t-1}^{(j)}) d\x_t  \\
&=  \sum_{j=1}^M w_{t-1}^{(j)} \int g(\y_t,\x_t|\x_{t-1}^{(j)}) d\x_t  \\
&=  \sum_{j=1}^M w_{t-1}^{(j)}g(\y_t|\x_{t-1}^{(j)}).
\end{align}


\qed

\begin{lemma}\label{lemma_7pitt}
For any $h\in\{1,...,t-1\}$ we have that 
\begin{equation}
\E\Big[\zest_{t-h:t}|\mcA_{t-h-1} \Big]  = \sum_{m=1}^M w_{t-h-1}^{(m)}g(\y_{t-h:t}|\x_{t-h-1}^{(m)}).
\label{eq_lemma_7_pitt}
\end{equation}
\end{lemma}
\noindent\emph{Proof:} We follow a proof by induction. First, note that \eqref{eq_lemma_7_pitt} is true for $h=0$ due to Lemma \ref{lemma_6_pitt}. Then, we assume that \eqref{eq_lemma_7_pitt} holds for a given $h$ and we will prove that it then holds for $h+1$. Let us start developing the left-hand side of \eqref{eq_lemma_7_pitt} for $h+1$ by first noting that $\zest_{t-h-1:t}= \zest_{t-h:t} \zest_{t-h-1}$. Then,
\begin{align}
\E\Big[\zest_{t-h-1:t}|\mcA_{t-h-2} \Big]  &= \E \Bigg[ \E \Big[\zest_{t-h:t}|\mcA_{t-h-1}  \Big]\zest_{t-h-1}|\mcA_{t-h-2}  \Bigg]\\
 &= \E \Bigg[ \Big[ \sum_{m=1}^M g(\y_{t-h:t}|\x_{t-h-1}^{(m)})w_{t-h-1}^{(m)} \Big]\zest_{t-h-1}|\mcA_{t-h-2}  \Bigg]\\
 \end{align}
 where we have simply substituted Eq. \eqref{eq_lemma_7_pitt} that we assume to hold for $h$. Next,
 \begin{align}
 \E\Big[\zest_{t-h-1:t}|\mcA_{t-h-2} \Big]  &= \E \Bigg[  \Big[ \sum_{m=1}^M g(\y_{t-h:t}|\x_{t-h-1}^{(m)})\frac{\widetilde w_{t-h-1}^{(m)}}{\sum_{j=1}^M \widetilde w_{t-h-1}^{(j)}} \Big]\frac{1}{M}\sum_{j=1}^M \widetilde w_{t-h-1}^{(j)}|\mcA_{t-h-2}  \Bigg]\\
 &= \E \Bigg[  \frac{1}{M} \sum_{m=1}^M g(\y_{t-h:t}|\x_{t-h-1}^{(m)}) \frac{g(\y_{t-h-1}|\x_{t-h-1}^{(m)})\sum_{j=1}^M w_{t-h-2}^{(j)}f(\x_{t-h-1}^{(m)}|\x_{t-h-2}^{(j)}) }{\psi_{t-h-1}(\x_{t-h-1}^{(m)})} |\mcA_{t-h-2}  \Bigg]\\
 &= \frac{1}{M} \sum_{m=1}^M \E\Bigg[   g(\y_{t-h:t}|\x_{t-h-1}^{(m)}) \frac{g(\y_{t-h-1}|\x_{t-h-1}^{(m)})\sum_{j=1}^M w_{t-h-2}^{(j)}f(\x_{t-h-1}^{(m)}|\x_{t-h-2}^{(j)}) }{\psi(\x_{t-h-1}^{(m)})} |\mcA_{t-h-2}  \Bigg]
 \end{align}
where we have substituted with the importance weights  $\widetilde w_{t-h-1}^{(m)}$ of Eq. 7 of the manuscript. Since, given $\mcA_{t-h-2}$, the particles at time $t$ are conditionally independent with pdf $\psi_{t-h-1}(\x_{t-h-1})$, all $M$ expectations are identical: 
\begin{align}
  \E\Big[\zest_{t-h-1:t}|\mcA_{t-h-2} \Big]  &= \int g(\y_{t-h:t}|\x_{t-h-1}) \frac{g(\y_{t-h-1}|\x_{t-h-1})\sum_{j=1}^M w_{t-h-2}^{(j)}f(\x_{t-h-1}|\x_{t-h-2}^{(j)}) }{\psi(\x_{t-h-1})}\psi(\x_{t-h-1})d\x_{t-h-1}\\
  &= \int g(\y_{t-h:t}|\x_{t-h-1}) g(\y_{t-h-1}|\x_{t-h-1})\sum_{j=1}^M w_{t-h-2}^{(j)}f(\x_{t-h-1}|\x_{t-h-2}^{(j)})d\x_{t-h-1} \label{sketchystep1}\\
    &= \int  g(\y_{t-h-1:t}|\x_{t-h-1})\sum_{j=1}^M w_{t-h-2}^{(j)}f(\x_{t-h-1}|\x_{t-h-2}^{(j)})d\x_{t-h-1} \label{sketchystep2}
 \end{align}
Step \eqref{sketchystep1} to \eqref{sketchystep2} is justified since $\y_{t-h:t} \indep \y_{t-h-1} | \x_{t-h-1}$, so we can replace $g(\y_{t-h:t}|\x_{t-h-1})$ in \ref{sketchystep1} with $g(\y_{t-h:t}|\y_{t-h-1},\x_{t-h-1})$ and then  $g(\y_{t-h-1:t}|\x_{t-h-1}) = g(\y_{t-h:t}|\x_{t-h-1}) g(\y_{t-h-1}|\x_{t-h-1})$ follows by the chain rule. Next,
\begin{align}
&= \sum_{j=1}^M   w_{t-h-2}^{(j)} \int  g(\y_{t-h-1:t},\x_{t-h-1}|\x_{t-h-2}^{(j)})  d\x_{t-h-1}\\
&= \sum_{j=1}^M   w_{t-h-2}^{(j)}   g(\y_{t-h-1:t}|\x_{t-h-2}^{(j)}) \\
\end{align}
which is the right-hand side of \eqref{eq_lemma_7_pitt}. \qed

\begin{theorem}\label{th_unbiased_z}
The OAPF estimator of the normalizing constant is unbiased, i.e., $\E[\zest_{1:t}] = p(\y_{1:t})$.
\end{theorem}
\noindent\emph{Proof:} The unbiasedness is a consequence of Lemma 2  with $h=t-1$.\qed

Now we look at the variance of the normalizing constant estimators. First, we establish a superiority in performance (i.e., equal or less variance) of the OAPF importance weights. This result is also used below to prove the convergence of the estimators by standard results in particle filtering. 

Let us particularize importance weights in OAPF for the case with $K=M$ as
 \begin{equation}\label{eq_weights_supp}
            \widetilde{w}_{t}^{(m)} = \frac{g(\y_t |  \x_{t}^{(m)}) \sum_{i=1}^{M} w_{t-1}^{(i)} f(\x_{t}^{(m)} |  \x_{t-1}^{(i)}) }{\sum_{i=1}^{M} \lambda_{t}^{(i)} q_{t}^{(i)}(\x_{t}^{(m)} |  \bar{\mathbf{x}}_{t-1}^{(i)})}.
        \end{equation}
We also consider the generalized APF weights given by
 \begin{align}\label{eq_gen_apf_weights}
\widetilde{v}_{t}^{(m)} &= \frac{g(\y_t |  \x_{t}^{(m)})   w_{t-1}^{(m)} f(\x_{t}^{(m)} |  \x_{t-1}^{(m)}) }{ \lambda_{t}^{(m)} q_{t}^{(m)}(\x_{t}^{(m)} |  \bar{\mathbf{x}}_{t-1}^{(m)})}.
\end{align}
These are generalized in the sense that the concrete APF described in the main paper is obtained by setting $\lambda_{t}^{(m)} \propto w_{t-1}^{(m)} g(\y_t|\boldsymbol{\mu}_{t}^{(m)})$ and propagating particles with transition kernels $f(\cdot)$, thus our following discussion holds for any choice of $\lambda_{t}^{(m)}$.
\begin{lemma}\label{lemma_var}
The conditional variance of $\zest_t^{\text{OAPF}}$ using the OAPF weights in \eqref{eq_weights_supp} is always less or equal than the same estimator $\zest_t^{\text{APF}}$ using the APF weights in  \eqref{eq_gen_apf_weights}.
\end{lemma}
 \noindent\emph{Proof:} First, note that $\widetilde{v}_{t}^{(m)}$ can be interpreted as an importance weight in an extended space on $\x_t$ and the auxiliary variable $m$ (see for instance \citep[Section 3.1]{Klaas05}  and \citep{pitt1999filtering,godsill2019particle}). Next, $\widetilde{w}_{t}^{(m)}$ can be interpreted as a version of $\widetilde{v}_{t}^{(m)}$ where both in the numerator (approximate filtering pdf) and denominator (proposal pdf), the auxiliary variable has been marginalized. Then, the variance inequality for each importance weight holds from the application of the variance decomposition lemma (also known as law of total variance). This proof generalizes the result in \citep{Klaas05} for any set of mixture weights $\{ \lambda_t^{(m)}\}_{m=1}^M$, with $\sum_{j=1}^M\lambda_t^{(j)}$ and $\lambda_t^{(m)}\geq 0$, for all $m$. Finally, since both $\hat Z_t^{\text{OAPF}}$ and $\hat Z_t^{\text{APF}}$ are constructed as the average of the OAPF and APF weights, respectively, the conditional variance of $\hat Z_t^{\text{OAPF}}$ is necessarily upper-bounded by that of $\hat Z_t^{\text{APF}}$. \qed
 
 We now address the consistency of the normalizing constant, $\zest_{1:t}$, and the self-normalized IS (SNIS) estimator $\widehat I(h_{t}) = \sum_{m=1}^M w_t^{(m)}h_t(\x_t^{(m)})$.
 \begin{corollary}\label{asd}
The OAPF estimator of the normalizing constant $\zest_{1:t}$ and the SNIS estimator $\widehat I(h_{t})$ are consistent, i.e., $\lim_{M\to\infty} \zest_{1:t} = p(\y_{1:t})$ and $\lim_{M\to\infty} \widehat I(h_{t}) = I(h_{t})$ a.s. (almost surely) for a finite $t$.
\end{corollary}
 \noindent\emph{Proof:} 
 The consistency of $\zest_{1:t}$ is a consequence of its unbiasedness, proved in Theorem \ref{th_unbiased_z}, and the variance inequality in Lemma \ref{lemma_var}, which ensures the variance convergence to zero a.s. when $N\to\infty$ since the APF, which upper-bounds its variance, is also consistent \citep[Section 3.6]{doucet2009tutorial}. A similar argumentation can be done for the SNIS estimator  $\widehat I(h_{t})$. Note that the SNIS estimator can be re-expressed as $\widehat I(h_{t}) = \sum_{m=1}^M \frac{\widetilde w_t^{(m)}}{M\widehat{Z_t}}h_t(\x_t^{(m)}) = \frac{1}{M}\sum_{m=1}^M \frac{\widetilde w_t^{(m)}}{\widehat{Z_t}}h_t(\x_t^{(m)}) $. Since $\zest_{t}$ is a consistent estimator of $p(\y_{t}|\y_{1:t-1})$, the denominator converges to $p(\y_{t}|\y_{1:t-1})$ while the numerator converges to $p(\y_{t}|\y_{1:t-1})I(h_t)$, when $N\to\infty$. Therefore, the ratio converges to $I(h_t)$ a.s. 
 \qed

\subsection{Additional Experiments and Results}
\subsubsection{Experiment 1}
We provide all necessary parameters to reproduce Figure 1 in the main paper. We recall that in this toy example we do the Bayesian recursion from $t-1$ to $t$ with $M=4$ particles. In Figure 1(a), we have set the particles $ \{  \bar{\x}_{t-1}^{(m)}  \}_{m=1}^{M=4} =  \{ 2, 2.5 , 3,3.5  \} $, the normalized weights $ \{ 3/10,3/10,1/5,1/5   \}$, likelihood centered at $3$, and $\sigma_{\text{lik}} = 0.8$, and $\sigma_{\text{kern}} = 0.5$.\\
In Figure 1(b), $ \{  \bar{\x}_{t-1}^{(m)}  \}_{m=1}^{M=4} =  \{ 2, 2.5, 5, 5.5  \} $, the normalized weights are $ \{ 7/22, 1/11, 1/2, 1/11   \}$, the likelihood is centered at $3.5$, and  $\sigma_{\text{lik}} = 1.2$, and $\sigma_{\text{kern}} = 0.5$. The proposals of all algorithms are then calculated as:
\begin{equation}\label{proposals}
    \sum_{m=1}^{4} \lambda_{t}^{(m)} f(\x_t | \overline{\x}_{t-1}^{(m)}) ,
\end{equation}
where the mixture weights  $\lambda_{t}^{(m)}$ for BPF are $w_{t-1}^{(m)}$, for APF are $ \propto w_{t-1}^{(m)} g(\y_t | \boldsymbol{\mu}_{t}^{(m)})$, for IAPF $\propto  g(\y_{t} | \boldsymbol{\mu}_{t}^{(m)}) \sum_{m=1}^{M} w_{t-1}^{(m)}  f( \boldsymbol{\mu}_{t}^{(m)} | \overline{\x}_{t-1}^{(m)}) / \sum_{m=1}^{M}  f( \boldsymbol{\mu}_{t}^{(m)} | \overline{\x}_{t-1}^{(m)}) $ and finally for OAPF they are the solution to the NNLS optimization problem. \\ 
As specified in the main paper and can be seen from \eqref{proposals}, we used transition kernels as proposal kernels for OAPF. Moreover, we used the centers of the transition kernels $\boldsymbol{\mu}_{t}^{(m)}$ as evaluation points, which in this case they correspond to the resampled particles $\{  \bar{\x}_{t-1}^{(m)}  \}_{m=1}^{M=4}$.

\subsubsection{Experiment 2}
In this Section we provide results for estimation of the marginal likelihood, additional results in the estimation of the posterior mean and relevant equations for the linear dynamical model (LDM) (Experiment 2 in the main paper). The model is given by
\begin{align}
    p(\x_0) &= \mathcal{N}_{\x_{0}}(\mathbf{m}_{0}, \boldsymbol{\Sigma}_{0}) \\
f\left(\mathbf{x}_{t} \mid \mathbf{x}_{t-1}\right) &=\mathcal{N}_{\mathbf{x}_{t}}\left(\mathbf{A x}_{t-1}+\mathbf{c}, \mathbf{R}\right) \\
g\left(\mathbf{y}_{t} \mid \mathbf{x}_{t}\right) &=\mathcal{N}_{\mathbf{y}_{t}}\left(\mathbf{C x}_{t}+\mathbf{g}, \mathbf{Q}\right) .
\end{align}
The posterior filtering distribution can be computed in closed form via the Kalman filter:
\begin{align}
    p(\mathbf{x}_t \mid \mathbf{y}_{1:t}) &= \mathcal{N}_{\mathbf{x}_t}(\boldsymbol{\mu}_{t}, \boldsymbol{\Sigma}_t) \\
    \boldsymbol{\mu}_t &= \overline{\boldsymbol{\mu}}_{t} + \mathbf{K} \left ( \mathbf{y}_t - \mathbf{C}\overline{\boldsymbol{\mu}}_{t} - \mathbf{g} \right )\\
    \boldsymbol{\Sigma}_t &= \left ( \mathbf{I} - 
    \mathbf{K} \mathbf{C} \right ) \overline{\boldsymbol{\Sigma}}_t  \\
    \overline{\boldsymbol{\mu}}_t &= \mathbf{A}\boldsymbol{\mu}_{t-1} + \mathbf{c} \\
    \overline{\boldsymbol{\Sigma}}_t &=  \mathbf{A} \boldsymbol{\Sigma}_{t-1} \mathbf{A}^\top + \mathbf{R} \\
    \mathbf{K} &= \overline{\boldsymbol{\Sigma}}_t \mathbf{C}^\top \left ( \mathbf{C} \overline{\boldsymbol{\Sigma}}_t \mathbf{C}^\top + \mathbf{Q} \right )^{-1} .
\end{align}
Moreover, $p(\y_{1:t})$ can also be computed in closed form from $p(\y_{t}|\y_{1:t-1})$. For numerical stability, one computes $\log p(\y_{1:t})$ and $\log p(\y_{t}|\y_{1:t-1})$, which are given by:
\begin{align}
    \log p(\y_{1:t}) &= \log p(\y_1) + \sum_{\tau = 2}^{t} \log p(\y_{\tau} | \y_{1:\tau -1}) \\
    \log p(\y_{\tau} | \y_{1:\tau -1}) &= -\frac{1}{2}\left[\log (|\mathbf{C}\overline{\boldsymbol{\Sigma}}_{\tau} \mathbf{C}^{\top} + \mathbf{Q}|)+(\mathbf{y}_{\tau}-\mathbf{C}\overline{\boldsymbol{\mu}}_{\tau}  - \mathbf{g})^{\mathrm{T}} (\mathbf{C}\overline{\boldsymbol{\Sigma}}_{\tau} \mathbf{C}^{\top} + \mathbf{Q})^{-1}(\mathbf{y}_{\tau}- \mathbf{C}\overline{\boldsymbol{\mu}}_{\tau}  - \mathbf{g})+ d_{\y_{\tau}} \ln (2 \pi)\right] \\
    \log p(\y_1 ) &= -\frac{1}{2}\left[\log (|\mathbf{C}\boldsymbol{\Sigma}_{0} \mathbf{C}^\top + \mathbf{Q} |)+(\mathbf{y}_{\tau}- \mathbf{C}\overline{\boldsymbol{\mu}}_{1}  - \mathbf{g} )^{\mathrm{T}} (\mathbf{C}\boldsymbol{\Sigma}_{0} \mathbf{C}^\top + \mathbf{Q} )^{-1}(\mathbf{y}_{\tau}- \mathbf{C}\overline{\boldsymbol{\mu}}_{1}  - \mathbf{g})+ d_{\y_{1}} \ln (2 \pi)\right] .
\end{align}

\begin{figure*}[t]
    \centering
    \begin{subfigure}[t]{.5\columnwidth}
    \centering
        \makebox[\textwidth]{\includegraphics[width=\columnwidth]{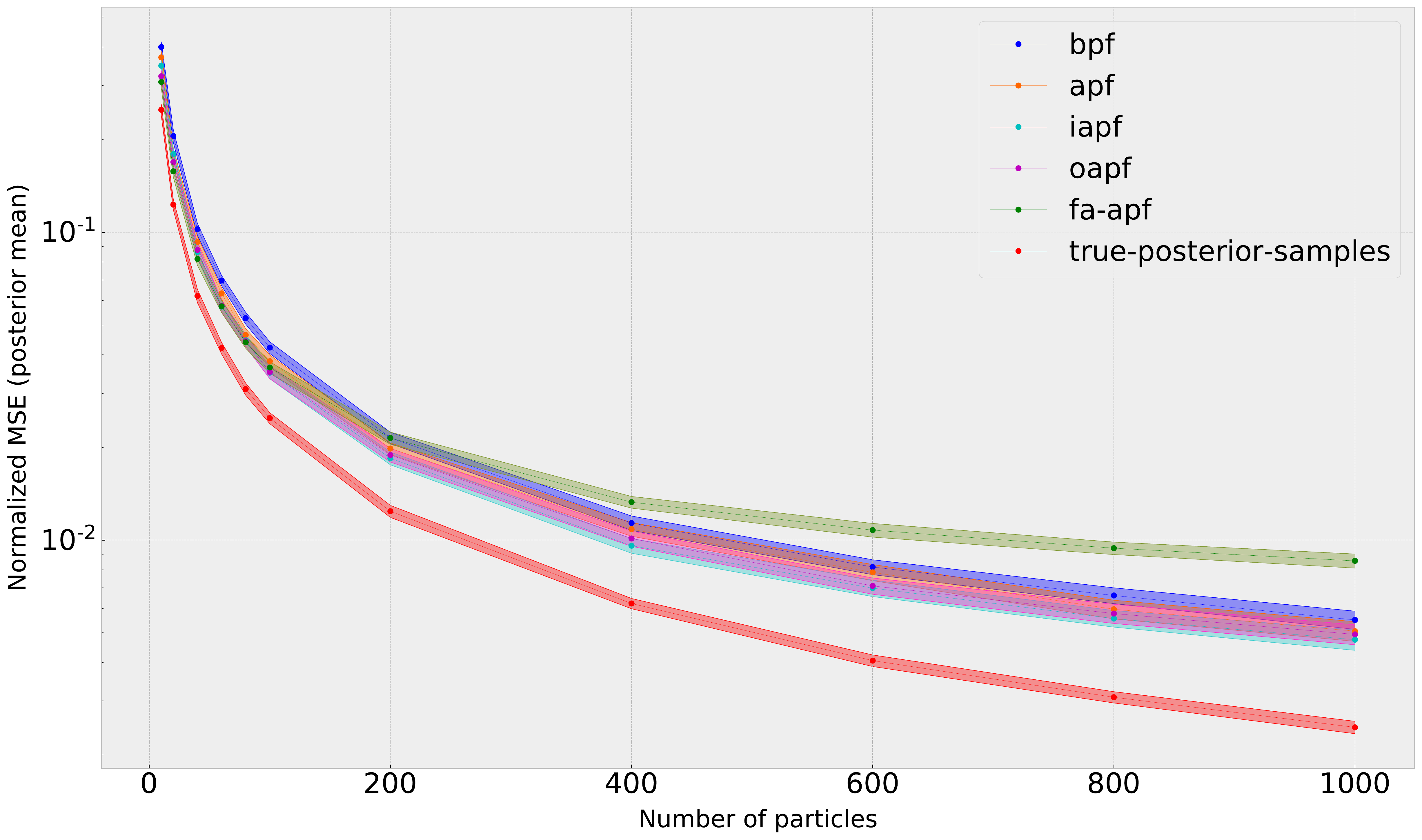}}
        \caption{ $d_{\x} = 2$, transition  covariance $2.5\mathbf{I}$, observation covariance $5\mathbf{I}$}
          \label{fig:1}
    \end{subfigure}%
    \begin{subfigure}[t]{.5\columnwidth}
    \centering
        \makebox[\textwidth]{\includegraphics[width=\columnwidth]{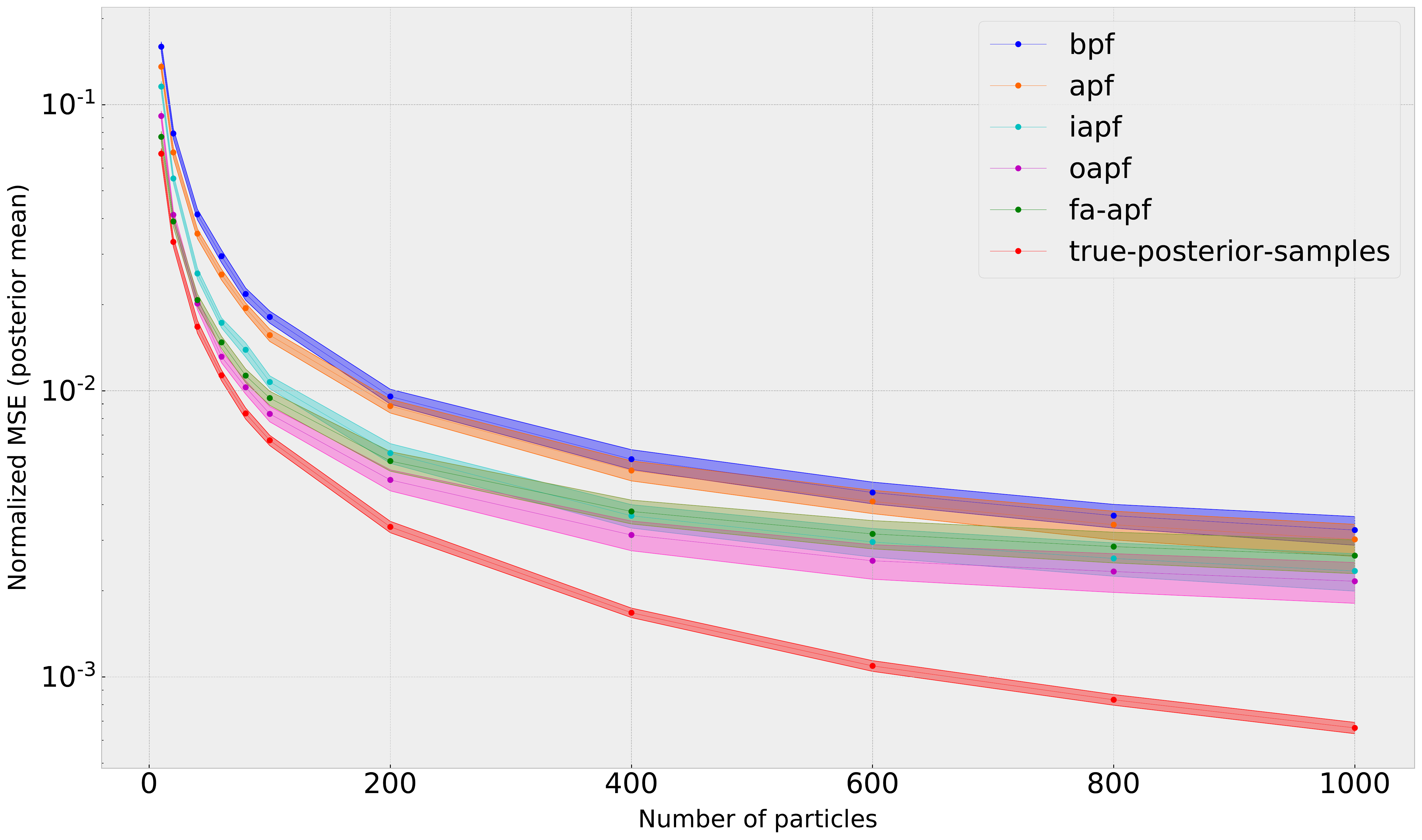}}
        \caption{$d_{\x} = 2$, transition  covariance $5\mathbf{I}$, observation covariance $2.5\mathbf{I}$}
          \label{fig:2}
    \end{subfigure}

    \begin{subfigure}[t]{.5\columnwidth}
    \centering
        \makebox[\textwidth]{\includegraphics[width=\columnwidth]{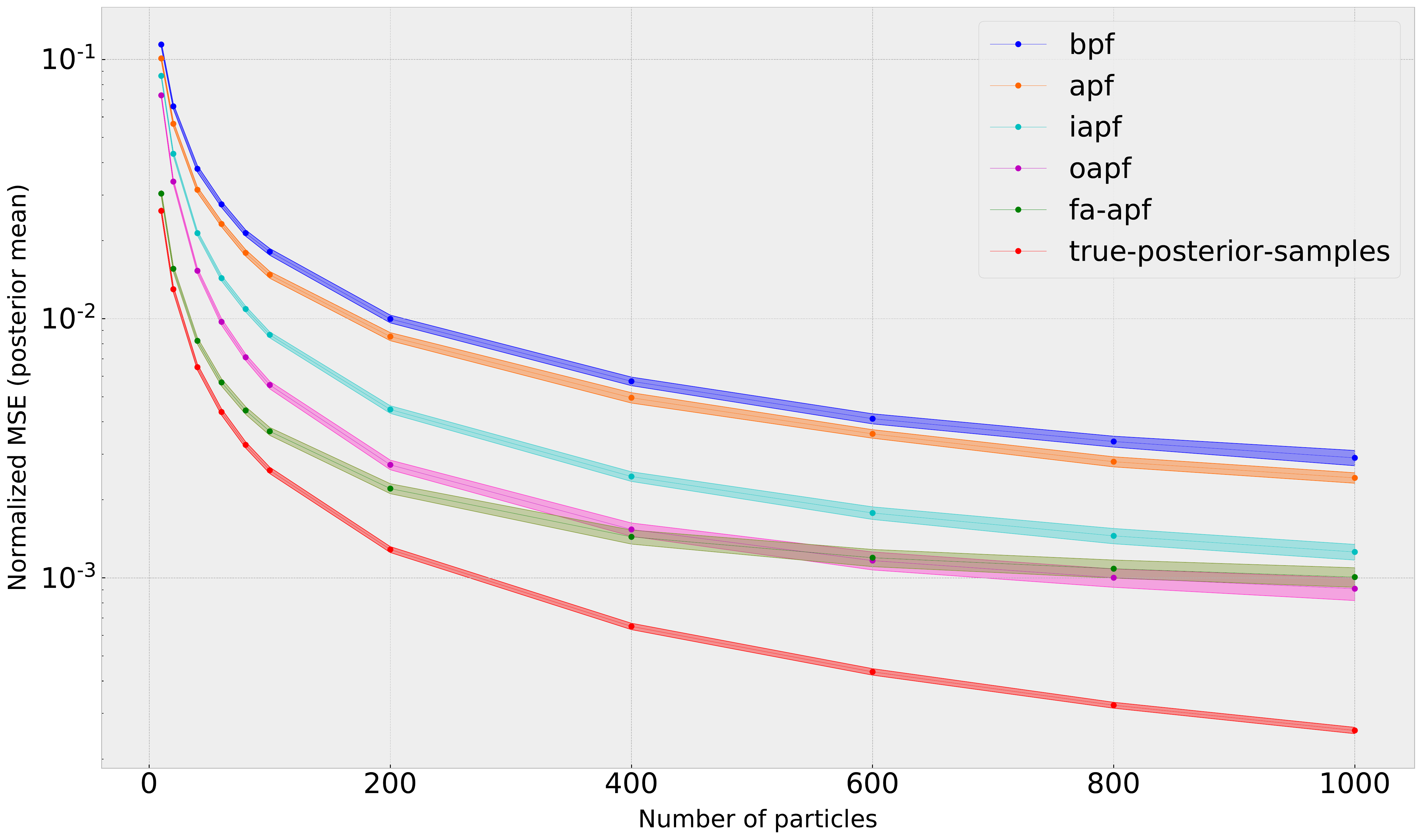}}
        \caption{$d_{\x} = 5$, transition  covariance $5\mathbf{I}$, observation covariance $2.5\mathbf{I}$}
          \label{fig:3}
    \end{subfigure}%
    \begin{subfigure}[t]{.5\columnwidth}
    \centering
        \makebox[\textwidth]{\includegraphics[width=\columnwidth]{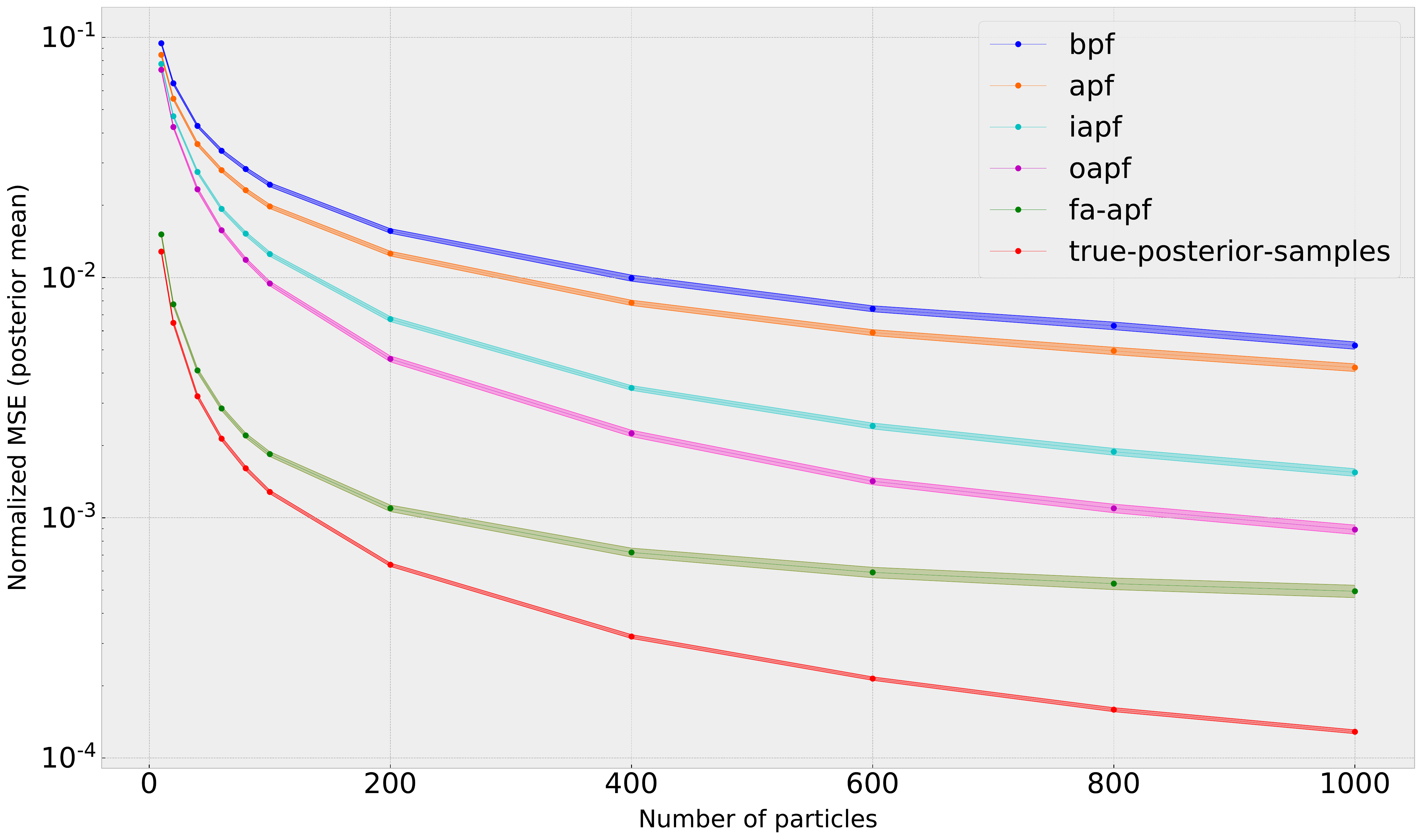}}
        \caption{$d_{\x} = 10$, transition  covariance $5\mathbf{I}$, observation covariance $2.5\mathbf{I}$}
          \label{fig:4}
    \end{subfigure}

\caption{ Here, we show results for the linear Gaussian model for additional settings (complement to Figure 2(a)). In all settings, OAPF ran with $\boldsymbol{K=5, E=5}$. Notice that the closed form FA-APF is often outperformed by OAPF, and sometimes even by BPF. }
\label{fig:final}
\end{figure*}

We set $\mathbf{R} = 5 \mathbf{I}$ and $\mathbf{Q}=  2.5 \mathbf{I}$. This setting is of particular interest, as the kernels overlap and the observations are very informative. Therefore, the setting is particularly advantageous for IAPF, and hence it is more difficult to beat its performance. Moreover,  $\mathbf{A}=\frac{1}{2}\mathbf{I}$ and $\mathbf{C}= \frac{1}{2}\mathbf{I}$. For $d_{\x} = 2$, then $\mathbf{c} = \mathbf{g}= (-2,2)^{\top} $ ; for  $d_{\x} = 5$, then $\mathbf{c} = \mathbf{g}= (-2,2,-2,2,-2)^{\top} $; similarly defined for $d_{\x} = 10$. The results for all $d_{\x}$ in the estimation of the marginal likelihood are shown in Table \ref{tab:lingauss}. We recall that OAPF ran with $\boldsymbol{K=5, E=5}$. This implies large computational savings with respect to the IAPF (or similarly to any other algorithm which uses the full mixture in the denominator of the importance weights). Finally, we also show additional results in the estimation of the posterior mean in Figure \ref{fig:final}.

\begin{table*}[h]
\caption{Additional results for Experiment 2 in main paper. The Table shows normalized MSE to the true marginal likelihood $p(\y_{1:t})$, with standard errors over 100 Monte Carlo runs. Recall that whenever $d_{\x} \in \{ 2,5 \}$ then $M=100$ and when $d_{\x} = 10$ then $M=1000$.}  \label{tab:lingauss}
\begin{center}
 \begin{tabular}{cccc}
Method  &  $d_{\x} = 2$ & $d_{\x} = 5$ & $d_{\x} = 10$ \\
\hline \\
 BPF   & $3.19 \cdot 10^{-7} \pm 3.14 \cdot 10^{-8}$   & 5.09 $\cdot 10^{-7}$ $\pm$ 4.40 $\cdot 10^{-8}$  & $1.539 \cdot 10^{-7} \pm 1.267 \cdot 10^{-8}$ \\
 APF          & $3.51 \cdot 10^{-7} \pm 3.88 \cdot 10^{-8}$   & 4.68 $\cdot 10^{-7}$ $\pm$ 4.49 $\cdot 10^{-8}$ & $1.333 \cdot 10^{-7} \pm 1.224 \cdot 10^{-8}$  \\
 IAPF             & $2.15 \cdot 10^{-7} \pm 2.25 \cdot 10^{-8}$ & 1.63 $\cdot 10^{-7}$  $\pm$ 1.58 $\cdot 10^{-8}$ & $6.330 \cdot 10^{-8} \pm  6.204 \cdot 10^{-9}$   \\
 OAPF              & $\boldsymbol{1.35 \cdot 10^{-7} \pm 1.23 \cdot 10^{-8}}$  & $ \boldsymbol{9.67 \cdot 10^{-8} \pm  9.03 \cdot 10^{-9}}$ & $ \boldsymbol{4.771 \cdot 10^{-8} \pm  5.329 \cdot 10^{-9}}$ \
\end{tabular}
\end{center}
\end{table*}

\subsubsection{Experiment 3}
In this experiment, we have used the standard parameters for the Lorenz model given by $(s,r,b) = (10,28,2.667)$. We set transition and observation noise as independent standard normally distributed random variables. In Figure \ref{fig:lorenzraj},  we show visually, as stated in the main paper, how a small change in $\Delta t$ can lead to very different trajectories of $\x_t$. The sensitivity to $\Delta t$, to the initialization, and even to the  parameters $(s,r,b)$, jointly with the strong non-linearity of the generated trajectories, make the Lorenz model particularly challenging.

\subsubsection{Experiment 4}
In this experiment, we have used a challenging multivariate stochastic volatility model, which is common in related works (see for instance  \citep{Guarniero2017-pz}). Additional results with parameters
$\mathbf{m} = \mathbf{0}, \mathbf{U}_0 = \mathbf{I}, \mathbf{U} =  \mathbf{I}, \mathbf{\phi} = \mathbf{1} $ are shown in Table \ref{tab:stochvol}.

\begin{table}[t]
\caption{ Results with additional parameters for Experiment 4. Note that when $d_{\x}=10$ then $M=1000$.  Averaged ESS and standard errors over 100 Monte Carlo runs.} \label{tab:stochvol} 
\begin{center}
  \begin{tabular}{cccc}
Method  &  $d_{\x} = 2$ & $d_{\x} = 5$ & $d_{\x} = 10$\\
\hline \\
BPF        & $50.8 \pm 0.2 $& $21.2 \pm 0.4$  & 46.6 $\pm$ 0.5 \\
APF          & $59.7 \pm 0.2 $ & $31.9 \pm 0.4$ &   83.9 $\pm$ 0.6  \\
IAPF             & $80.5 \pm 0.1$ & $49.4 \pm 0.5$ & $199.9 \pm 1.7 $\\
OAPF             &  $\boldsymbol{92.6 \pm 0.1 }$ & $\boldsymbol{59.5 \pm 0.7} $& $\boldsymbol{229.5 \pm 2.4}$
\end{tabular}
\end{center}

\end{table}

\section{Discussion  on number of evaluation points and kernels}
\begin{figure*}[t]
    \centering
     \makebox[\textwidth]{\includegraphics[width=0.5\columnwidth]{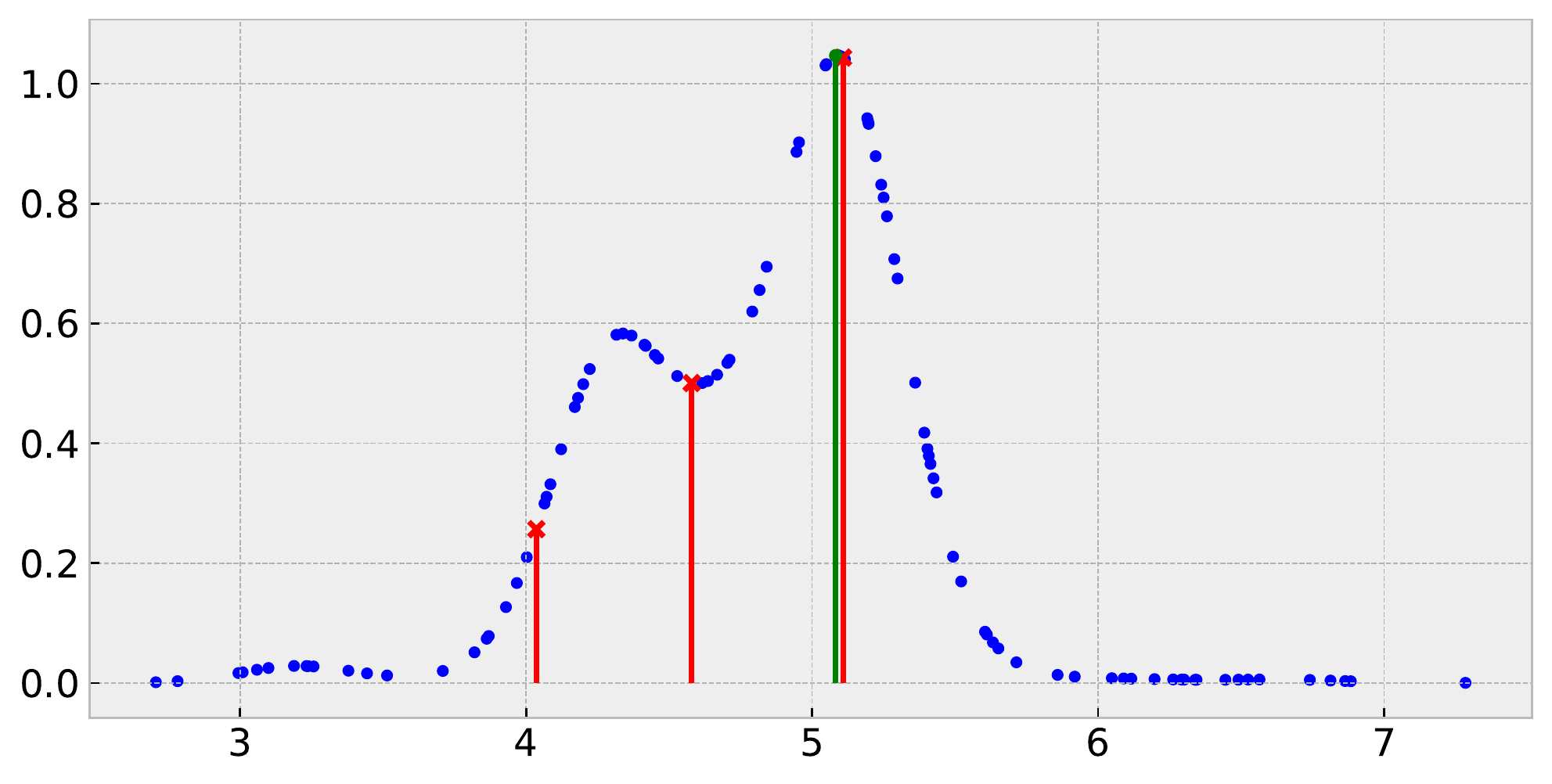}}
\caption{ In blue, evaluations of a posterior generated by multiplying a mixture of 100 Gaussian kernels with a Gaussian likelihood. Building the OAPF proposal would take approximately $100^3$ computation steps. However, one can see that it seems highly wasteful to adapt the mixture of Gaussian kernels proposal by evaluating at \emph{all blue points}: e.g., if the proposal matches the posterior at the rightmost red point, it will probably go through the green point too: we do not need to include that point in the optimization. Moreover, it may also be wasteful to match proposal and target at points where the target has little probability mass. From these considerations, one may consider that trying to only match the three highlighted red points (and perhaps a couple more), would likely result in a proposal that is closely as good as the one we would get by using all blue points. }
\label{fig:evalpoints}
\end{figure*}

\begin{figure*}[t]
    \centering
    \begin{subfigure}[t]{.5\columnwidth}
    \centering
        \makebox[\textwidth]{\includegraphics[width=\columnwidth]{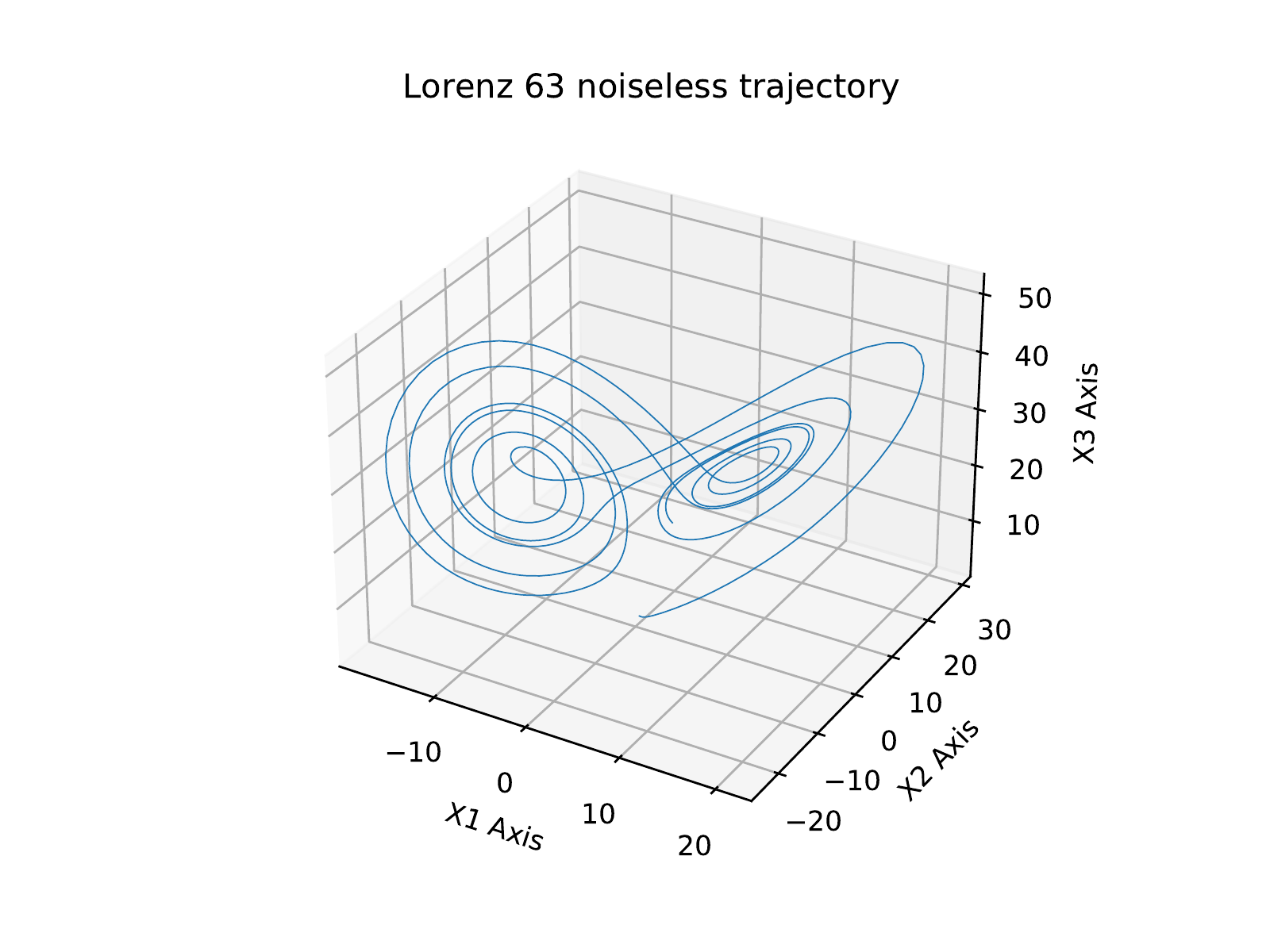}}
        \caption{ }
          \label{fig:lorenz1}
    \end{subfigure} \hfill \begin{subfigure}[t]{.5\columnwidth}
    \centering
        \makebox[\textwidth]{\includegraphics[width=\columnwidth]{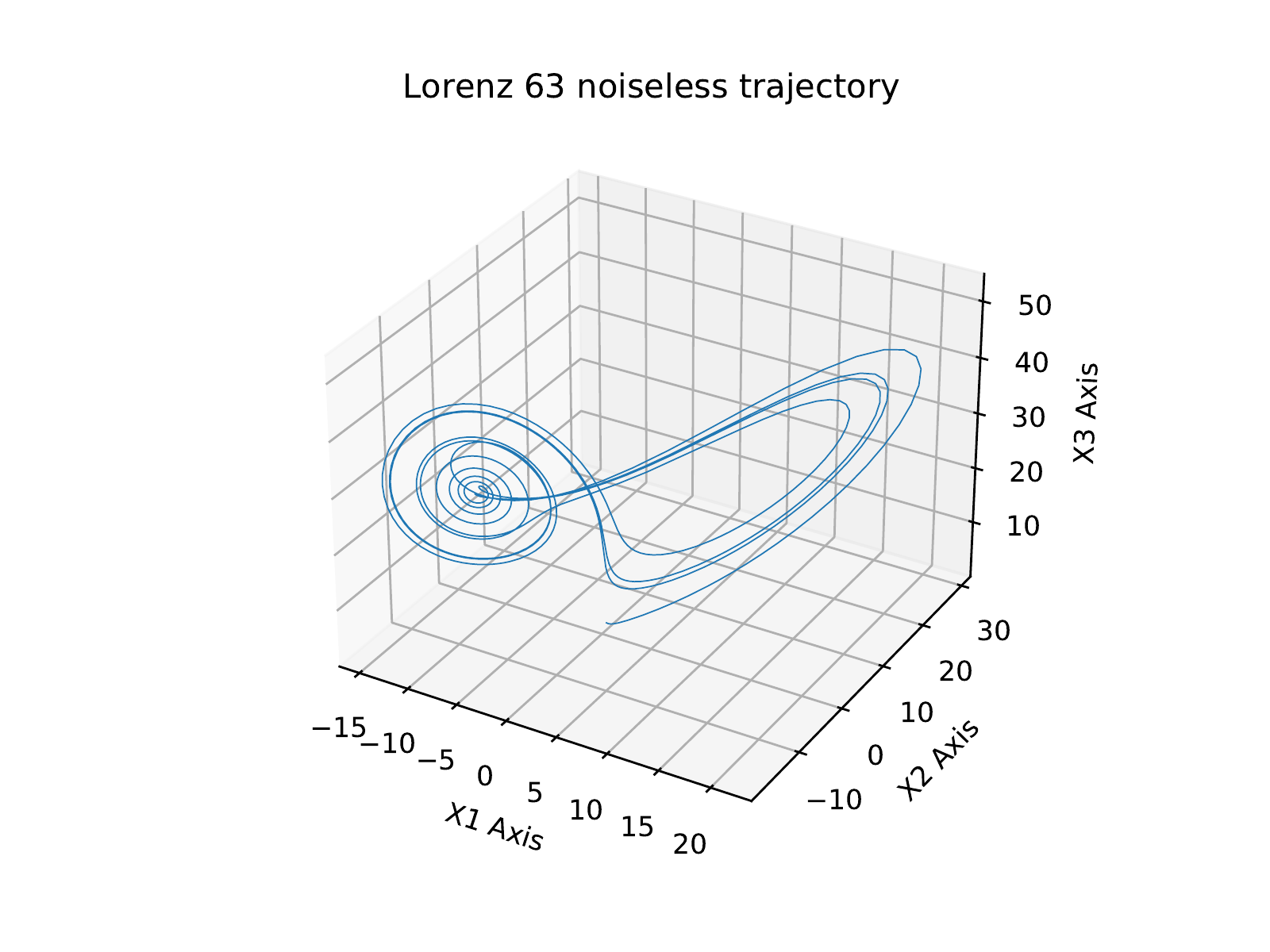}}
        \caption{ }
          \label{fig:lorenz2}
    \end{subfigure}
\caption{In this figure, we show the noiseless versions of a trajectory for $\Delta t = 0.01$ (a) and $\Delta t = 0.008$ (b) to emphasize how different trajectories can be in a Lorenz 63 model even with small parameter changes. Note that particle filters will have to deal with both transition noise and observation noise. }
\label{fig:lorenzraj}
\end{figure*}

We expand here our intuition for the need of only few kernels/evaluation points in many scenarios (see Figure \ref{fig:evalpoints} for explanation).
\section{Discussion on the fully adapted PF}
Previous works have discussed how the FA-APF described in \citep{pitt1999filtering,pitt2012some} optimality criterion is not optimal in a global sense: the main intuition they provide is that it only minimizes the one-step variance of the importance weights \citep{johansen2008note,whiteley2011auxiliary,chopin2020introduction}. Here, we will provide a perspective inspired by MIS to informally explain how FA-APF can be suboptimal in general. \\
Let us assume that we have access to $M$ samples simulated exactly from the filtering distribution at time $t-1$:
\begin{equation}
    \x_{t-1}^{(m)} \sim p(\x_{t-1} |  \y_{1:t-1}) \qquad m=1, \dots, M  .
\end{equation}
These $M$ samples can be used to form a particle approximation of $p(\x_{t-1} |  \y_{1:t-1})$ simply as:
\begin{equation}\label{eq:appproxpred}
    p(\x_{t-1} |  \y_{1:t-1}) \approx \frac{1}{M} \sum_{m=1}^{M} \delta_{\x_{t-1}^{(m)}} .
\end{equation}
This particle approximation can in turn be used to approximate the intractable integral in the definition of the filtering posterior and form an approximation to it:
\begin{align}
    p(\x_t | \y_{1:t}) &\propto g(\y_t | \x_t ) \int f(\x_t | \x_{t-1}) p(\x_{t-1} | \y_{1:t-1}) \mathrm{d}\x_{t-1} \label{eq:unnormalizedpost}\\ 
    & \approx  g(\y_t | \x_t ) \int f(\x_t | \x_{t-1})   \frac{1}{M} \sum_{m=1}^{M} \delta_{\x_{t-1}^{(m)}} \mathrm{d}\x_{t-1} \qquad \text{substituting}~ p(\x_{t-1} | \y_{1:t-1}) ~  \text{for approximation in Eq. \eqref{eq:appproxpred}} \\
    &= g(\y_t | \x_t ) \frac{1}{M}\sum_{m=1}^{M} f(\x_{t} | \x_{t-1}^{(m)}) \label{eq:approx_particle} 
\end{align}
Now, we will exploit the identity used by the FA-APF. The identity in question is:
\begin{align}\label{eq:fapfeq}
    p(\x_t |  \x_{t-1}, \y_{t}) = \frac{g(\y_t | \x_{t}) f(\x_t | \x_{t-1})}{p(\y_t | \x_{t-1})} ;
\end{align}
often the term $p(\y_t | \x_t-1)$ is referred to as \emph{predictive likelihood}. The FA-APF propagates each particle $m$ using $ p(\x_t |  \x_{t-1}^{(m)}, \y_{t})$ and resamples with weights  $\frac{p(\y_t | \x_{t-1}^{(m)})}{M}$ . It is easy to derive that this leads to constant importance weights $w_{t}^{(m)}$, when these are defined as :
\begin{equation}
w_{t}^{(m)} = \frac{p(\x^{(m)}_{1:t} | \y_{1:t})} {q(\x^{(m)}_{1:t-1} | \y_{1:t}) q(\x^{(m)}_t | \y_{t}, \x^{(m)}_{t-1}) }
\end{equation}
using a joint proposal and target as common in SMC. \\
Our observation is that the choices made by FA-APF can be viewed as sampling from the mixture in Eq. \eqref{eq:approx_particle}, when rearranged using \eqref{eq:fapfeq}:
\begin{equation}
   \frac{1}{M} g(\y_t | \x_t ) \sum_{m=1}^{M} f(\x_{t} | \x_{t-1}^{(m)})  = \frac{1}{M} \sum_{m=1}^{M} p(\y_t | \x_{t-1}^{(m)}) p(\x_{t} | \x_{t-1}^{(m)}, \y_t)  ,
\end{equation}
since $g(\y_t | \x_t ) f(\x_{t} | \x_{t-1}^{(m)}) = p(\x_{t} | \x_{t-1}^{(m)}, \y_t) p(\y_t | \x_{t-1}^{(m)})$, where indeed i.i.d. sampling from this mixture is equivalent to resampling and propagating in FA-APF.\\
This observation highlights some of the assumptions behind FA-APF from a different perspective: firstly, we assumed i.i.d. samples from the true filtering distribution at $t-1$ were available; secondly, we formed a particle approximation to $p(\x_t | \y_{1:t-1})$ in Eq.  \eqref{eq:unnormalizedpost} which may be more or less accurate depending on the situation. Therefore, the FA-APF choices of resampling weights and kernels, even when analytically available, can still lead to very poor performance if (1) the previous set of samples is a bad approximation of $p(\x_{t-1} | \y_{1:t-1})$, and consequently if the approximation to the predictive distribution $p(\x_{t} | \y_{1:t-1})$ is poor.

\clearpage

\bibliography{Branchini_500}

\end{document}